\documentclass[12pt]{article}
\usepackage{a4wide, amsmath, latexsym, epsfig, graphicx, rotating, fancyhdr, tabularx, afterpage}
\usepackage{cite}
\usepackage{axodraw}
\usepackage{subfigure}
\usepackage{color}
\usepackage{epstopdf}
\usepackage{ragged2e}
\begin{document}
\begin{titlepage}
\begin{flushright}
{\bf IFJPAN-IV-2017-6}
\end{flushright}
\vspace*{0.5cm}
\begin{center}
{\Large \bf Extra lepton pair emission corrections to Drell-Yan processes in
{\tt PHOTOS} and {\tt SANC}}
\end{center}

\vspace*{0.5cm}
\begin{center}
S.~Antropov$^a$, A.~Arbuzov$^{b,c}$, R.~Sadykov$^d$, Z.~Was$^{a,e}$.\\
\vspace*{0.5 cm}
{\small $^a$ Institute of Nuclear Physics, Polish Academy of Sciences, ul. Radzikowskiego 152, 31-342 Krakow, Poland;}\\
{\small $^b$ Bogoliubov Laboratory for Theoretical Physics, JINR, Dubna 141980, Russia;}\\
{\small $^c$ Dubna State University, Dubna 141982, Russia;}\\
{\small $^d$ Dzhelepov Laboratory of Nuclear Problems, JINR, Joliot-Curie str. 6, 141980 Dubna, Russia;}\\
{\small $^e$ Theory Group, Physics Department, CERN, CH-1211, Geneva 23, Switzerland.}
\end{center}

\date{\today}

\vspace*{0.5 cm}
\begin{center}
{\bf Abstract}
\end{center}

In the paper we present results for final state emissions of lepton pairs in decays of heavy
intermediate states such as Z boson. Short presentations of {\tt PHOTOS} and {\tt SANC}
algorithms and physics assumptions are given. Numerical distributions of relevance for LHC
observables are shown. They are used in discussions of systematic errors in the predictions of
pair emissions as implemented in the two programs. Suggestions for the future works are given.
Present results confirm, that for the precision of 0.3\% level, in simulation of final state the pair
emissions can be avoided. For the precision of 0.1-0.2\%, the results obtained with the presented
programs should be enough. To cross precision tag of 0.1\%, the further work is however required.
\vfill
{\small
\begin{flushleft}
{\bf {IFJPAN-IV-2017-6\\ June 2017}}
\end{flushleft}
}
\end{titlepage}
\section{Introduction}
With the increasing precision of measurements more detailed theoretical calculations are needed for
interpretation of results in the language of physics parameters such as masses or couplings of Z and
W bosons. In the present note, we
concentrate on effects and uncertainties related to emission of real lepton pair in association with
Drell-Yan processes. Our work is a direct continuation of \cite{Arbuzov:2012dx}, that is why we will
omit many definitions included in that paper. We will concentrate on the effects related to additional
pair emissions in decays of heavy bosons, mainly $Z$.

Our main goal is to study the effect of light pair emission $f\bar{f} $ in neutral current Drell-Yan
process $q\bar{q} \to \gamma/Z \to \ell^+\ell^-(f\bar{f})$ for $pp$ collisions at the LHC. We
consider the cases $\ell = e,\mu$ and $f = e,\mu$. This effect should be included starting from the
second order of QED, i.e. from the $\mathcal{O} (\alpha^2)$ corrections. The typical Feynman
diagrams for pair corrections are shown in Fig.~\ref{fig:diagpair}.
\begin{figure}[htp!]
{\sl Real pair emission}
\begin{center}
\begin{picture}(500,100)(0,0)
\ArrowLine(60,50)(30,80) \Text(20,80)[]{$\bar{q}$}
\ArrowLine(30,20)(60,50) \Text(20,20)[]{$q$}
\Vertex(60,50){2}
\Photon(60,50)(100,50){3}{7} \Text(80,60)[]{$\gamma/Z$}
\Vertex(100,50){2}
\ArrowLine(150,100)(100,50) \Text(160,100)[]{$\ell^+$}
\ArrowLine(125,25)(150,0) \Text(160,0)[]{$\ell^-$}
\ArrowLine(100,50)(125,25) \Text(105,35)[]{$\ell^-$}
\Photon(125,25)(165,25){3}{7} \Text(145,35)[]{$\gamma$}
\Vertex(125,25){2}
\Vertex(165,25){2}
\ArrowLine(190,45)(165,25) \Text(200,45)[]{$\bar{f}$}
\ArrowLine(165,25)(190,5) \Text(200,5)[]{$f$}

\ArrowLine(310,50)(280,80) \Text(270,80)[]{$\bar{q}$}
\ArrowLine(280,20)(310,50) \Text(270,20)[]{$q$}
\Vertex(310,50){2}
\Photon(310,50)(350,50){3}{7} \Text(330,60)[]{$\gamma/Z$}
\Vertex(350,50){2}
\ArrowLine(375,75)(350,50) \Text(358,68)[]{$\ell^+$}
\ArrowLine(400,100)(375,75) \Text(410,100)[]{$\ell^+$}
\ArrowLine(350,50)(400,0) \Text(410,0)[]{$\ell^-$}
\Photon(375,75)(415,75){3}{7} \Text(395,65)[]{$\gamma$}
\Vertex(375,75){2}
\Vertex(415,75){2}
\ArrowLine(440,95)(415,75) \Text(450,95)[]{$\bar{f}$}
\ArrowLine(415,75)(440,55) \Text(450,55)[]{$f$}
\end{picture}
\end{center}

{\sl Virtual pair correction}
\begin{center}
\begin{picture}(200,100)(0,0)
\ArrowLine(60,50)(10,100) \Text(0,100)[]{$\bar{q}$}
\ArrowLine(10,0)(60,50) \Text(0,0)[]{$q$}
\Vertex(60,50){2}
\Photon(60,50)(100,50){3}{7} \Text(80,60)[]{$\gamma/Z$}
\Vertex(100,50){2}
\Vertex(130,80){2}
\ArrowLine(130,80)(100,50) \Text(160,100)[]{$\ell^+$}
\ArrowLine(150,100)(130,80) \Text(108,68)[]{$\ell^+$}
\ArrowLine(130,20)(150,0) \Text(160,0)[]{$\ell^-$}
\ArrowLine(100,50)(130,20) \Text(105,35)[]{$\ell^-$}
\Vertex(130,20){2}
\ArrowArc(130,50)(10,90,270)
\ArrowArc(130,50)(10,270,90)
\Vertex(130,40){2}
\Vertex(130,60){2}
\Photon(130,60)(130,80){3}{4}
\Photon(130,40)(130,20){3}{4}
\Text(115,50)[]{$\bar{f}$}
\Text(145,50)[]{$f$}
\Text(140,70)[]{$\gamma$}
\Text(140,30)[]{$\gamma$}
\end{picture}
\end{center}
\caption[]{Feynman diagrams for real and virtual pair correction.
\label{fig:diagpair}}
\end{figure}

The {\tt PHOTOS} \cite{Barberio:1990ms,Barberio:1993qi,Golonka:2005pn,Nanava:2006vv,Golonka:2006tw,Nanava:2009vg,Davidson:2010ew} and {\tt SANC} \cite{Andonov:2004hi,Arbuzov:2005dd,Arbuzov:2007db,Arbuzov:2007kp,Andonov:2007zz, Andonov:2008ga,Andonov:2009nn,Bardin:2012jk,Arbuzov:1999cq,Arbuzov:2001rt}
Monte Carlo programs use different an approximations for the effect under study. We will show the program features important for effect of pair emissions respectively in Section $2$ and $3$. The numerical comparison of the results from the two programs and benchmark semi-analytical calculations follows. In Section $4$ the definition of our tests distributions is given. Main results are also collected in this section. Section 5 is devoted to the case of mixed pair and photon emissions and summary Section $6$ closes the paper. Extensive Appendix collects result of our new semi-analytical calculations for pair emissions which is used to obtain numerical results necessary to understand origin of {\tt PHOTOS}-{\tt SANC} differences.

\section{Pair corrections in {\tt PHOTOS}}
\label{sec:PHOTOSpairs}
The basis of {\tt PHOTOS} algorithm is of the after-burner type. For the previously generated event,
with a certain probability, a decay vertex can be replaced with the one featuring additional photons
(similar solution for additional lepton pairs is installed) \cite{Davidson:2010ew}.

For that purpose, {\tt PHOTOS} uses the exact phase space parametrizations.
The best description of its phase-space generation is given in \cite{Nanava:2009vg}.
Case of pair emission is quite analogous and the kinematical configuration for each decay is first
deconvoluted into angular parametrization of two body decay into {\tt emitter} and {\tt spectator}\footnote{ The spectator may represent multiple particles. But as corresponding Jacobians for phase space parametrization do not need to be modified we may omit details from our brief presentation.}.
The corresponding angles, together with extra generated ones, provide parametrization of four body
phase space; all necessary phase-space Jacobians are calculated and taken
into account. Corresponding algorithm for phase-space is  also exact in
the case of emission of additional lepton pairs.

It was checked with samples of 100 million events that once matrix element is set to unity, flat
four body phase space generation is achieved. This was checked with default test of {\tt MC-TESTER}
\cite{Davidson:2008ma}.

Before matrix element installation, pre-samplers were introduced and checked as well, respectively for
collinear, small virtuality and small energy of virtual photon enhancements. For the case of two
channels of singularity structure, two pre-samplers are needed. In this case phase space
parametrization remains exact. However, when further particles, such as additionally generated photons
appear, parametrization of phase-space ceases to be exact. This is due to the matching of Jacobians for
distinct generation branches. This non-exactness appear as in multi-photon's emission or in any
other case of more than two body decays in {\tt PHOTOS} operation.

The probability distribution for pair emission is independent from the
Born-level matrix-element squared.
 It is defined by  integrand for  $\tilde{B}_f$  (formula (1) from \cite{Jadach:1993wk}). Such a formula is valid for the soft
pairs emissions but is applied, at present, in {\tt PHOTOS} Monte Carlo algorithm over the entire phase space. If
the energy of the emitted pair is smaller than $\Delta$ ($2m_f \ll \Delta \ll \sqrt{s}$) then the
formula~(11) from \cite{Jadach:1993wk} is valid too. It was used to check the validity of {\tt PHOTOS}
prediction in the soft region. Agreement at the expected level of few percents of pair effect was found for
electrons and muons, and for several choices of maximum energy of emitted lepton pairs.

Further work on matrix element used in {\tt PHOTOS} can be continued, once tests of the present
version are completed. The corresponding task is going to be rather straightforward. The presently
used matrix element is calculated in separate program unit directly from the decay products
four-vectors. Test, with the help of {\tt KORALW} \cite{Jadach:1998gi} Monte Carlo featuring matrix element for $Z$ to four fermions decay, is reported.

Emission of pairs can be simultaneous with emission of real photons. The algorithm can be used in
such case as well. The solution is consistent for the leading logarithms with evolution
equations. Numerical tests were not performed because pair correction is too small to justify
the effort. It was only checked that the variants of algorithm do not lead to numerically
sizable effects.

For the virtual correction emulation, the sum rule is used.
\section{Pair corrections in {\tt SANC}} \label{sec:SANC}
In {\tt SANC} the leading logarithmic approximation (LLA) was applied to take into account the
corrections of the orders $\mathcal{O}(\alpha^n L^n)$, $n = 2, 3$. The contribution of pair emission
is approximated by the formula (8) from \cite{Arbuzov:2012dx}, where big logarithms
$ L(m_\ell,\mu) = \log{(\mu^2 / m_\ell^2)}$ depends on the lepton mass $m_\ell$ and on the
factorization scale $\mu$. For the sake of comparison we keep only the term proportional to
$\alpha^2$ in the above-mentioned formula, i.e. the following expression is used:
\begin{eqnarray}
{\mathcal D}^{\mathrm{pair}}_{\ell\ell}(y,L) =
\biggl(\frac{\alpha}{2\pi}(L-1)\biggr)^2
\biggl[\frac{1}{3}P^{(1)}(y) + \frac{1}{2}R^s(y)\biggr]
\label{eq:sancpair}
\end{eqnarray}

\section{Setup for comparison and numerical results}
For the comparison we used the same scheme and the values of input parameters as in
\cite{Arbuzov:2012dx} (eq.(2)). The cut on invariant mass $M(\ell^+\ell^-) > 50 \text{ GeV} $ was
imposed.

\begin{figure}[htp!]
\includegraphics[width=0.5\textwidth]{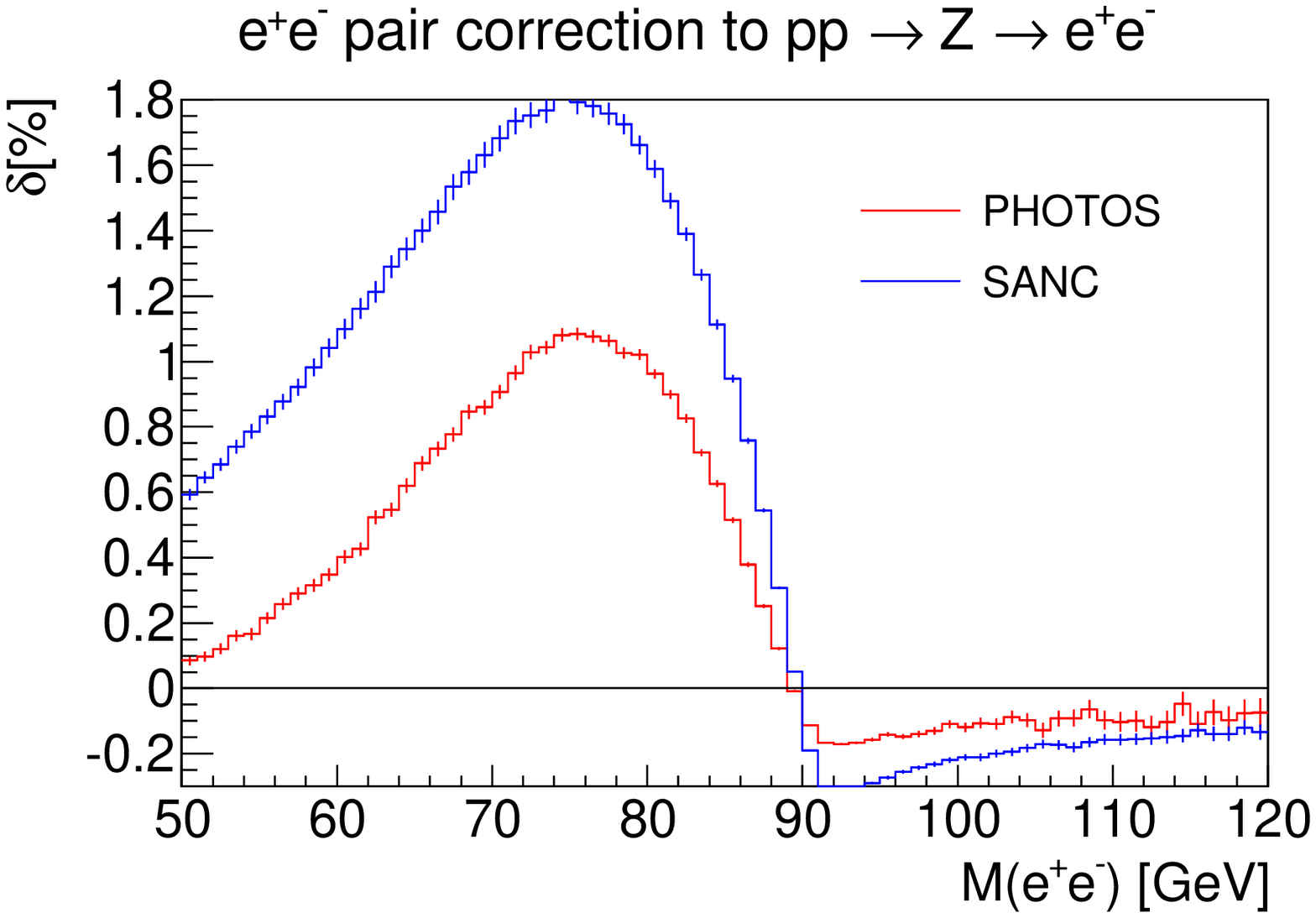}
\includegraphics[width=0.5\textwidth]{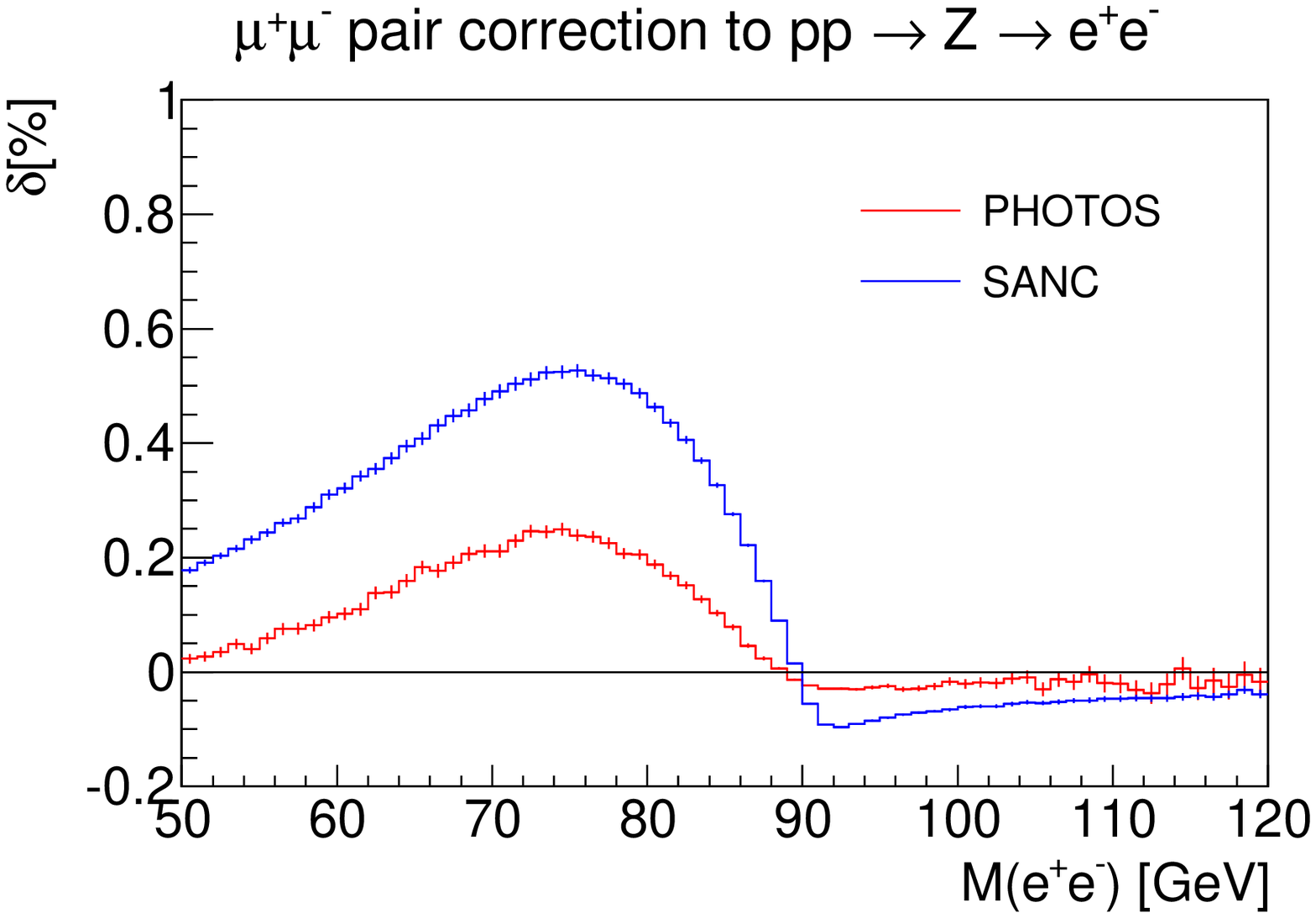}
\caption[] {Corrections $\delta$ in \% for invariant mass $M(e^{+}e^{-})$ distribution in
$Z \to e^+e^-$ decay due to extra $e^+e^-$ (left) or $\mu^+\mu^-$ (right) pair emission.
\label{fig:Zee}}
\end{figure}

\begin{figure}[htp!]
\includegraphics[width=0.5\textwidth]{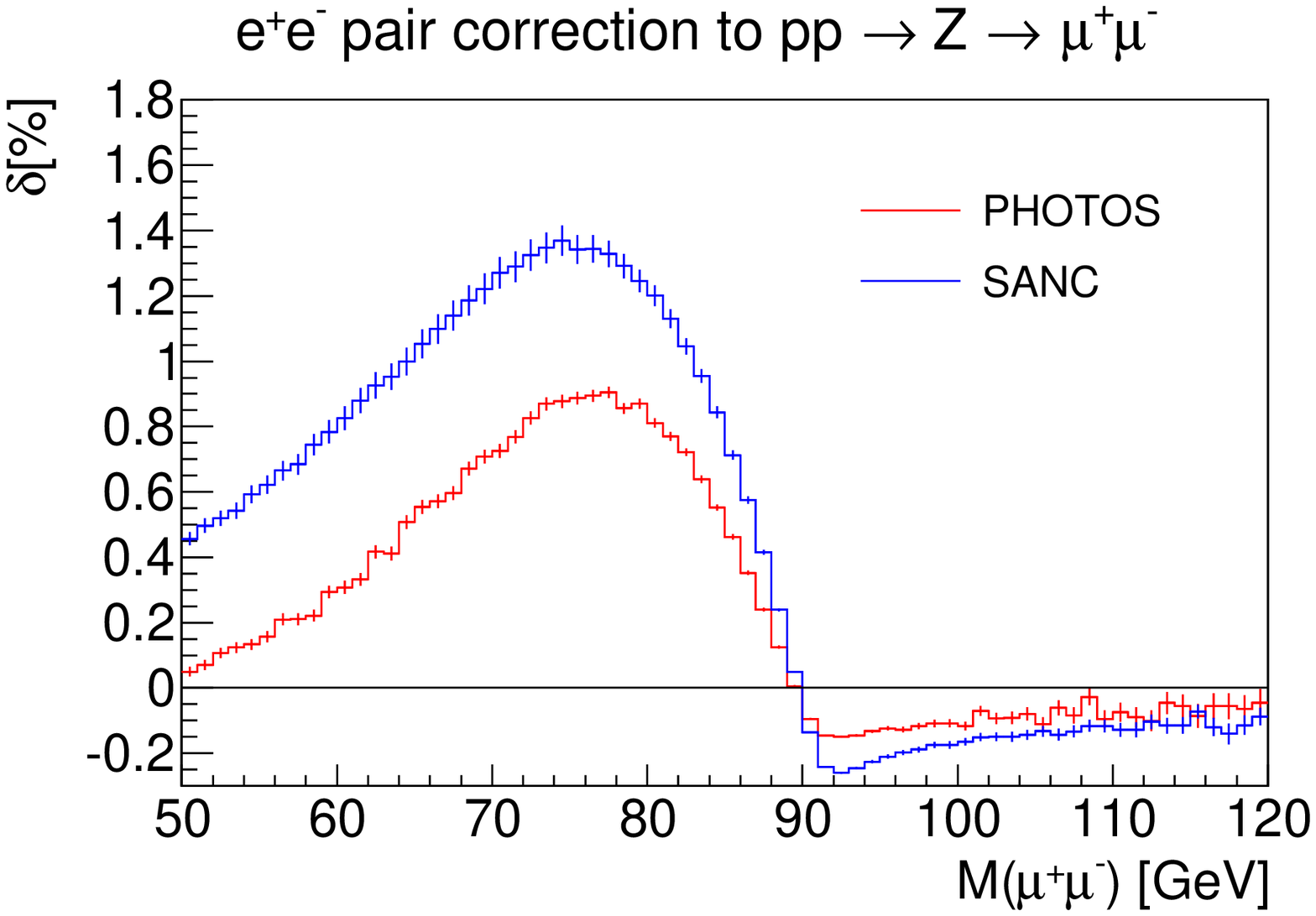}
\includegraphics[width=0.5\textwidth]{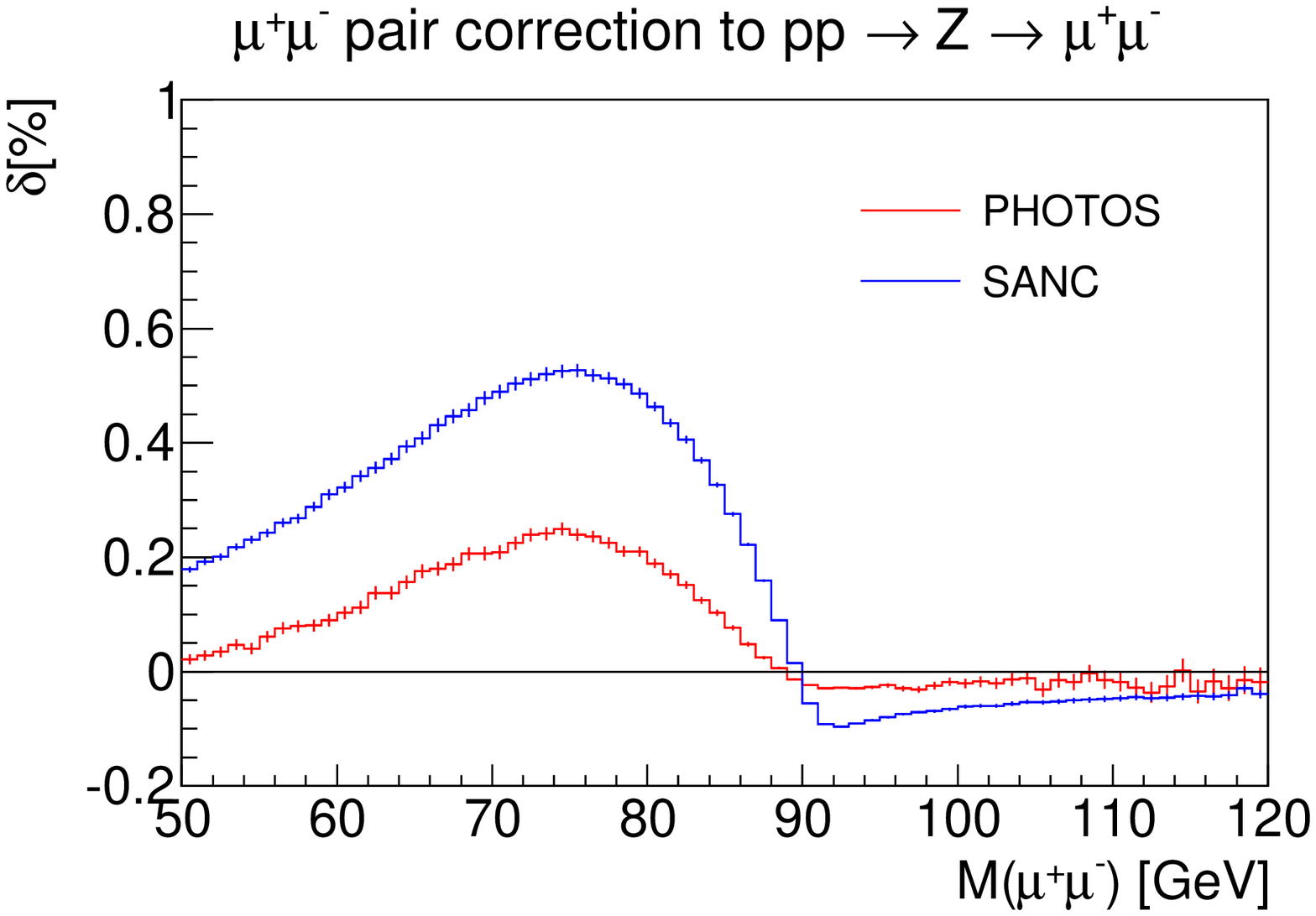}
\caption[] {Corrections $\delta$ in \% for invariant mass $M(\mu^{+}\mu^{-})$ distribution in
$Z \to \mu^+\mu^-$ decay due to extra $e^+e^-$ (left) or $\mu^+\mu^-$ (right) pair emission.
\label{fig:Zmm}}
\end{figure}

We define the correction as $\delta^{pair} = (\sigma^{pair}-\sigma^{Born})/\sigma^{Born}$. The
results for distribution of invariant mass $M(\ell^{+}\ell^{-})$ are presented in Fig.~\ref{fig:Zee}
and Fig.~\ref{fig:Zmm} for  {\tt PYTHIA} generated sample of Drell-Yan processes at $14$~TeV center of mass energy  pp collisions and final state of electron and muon pairs respectively.

An agreement between pair implementation with the help of {\tt PHOTOS} and
{\tt SANC} seems not to be sufficient, differences are  dominated, as we will see later,
by non leading terms and of rather hard pair emission. Let us continue with discussion of results.

The comparison between {\tt HORACE}~\cite{CarloniCalame:2003ux} and {\tt SANC} of pair contributions is presented in the Ref.~\cite{Alioli:2016fum}.
One can see, that a better agreement was found in this case, but the implementation of pair corrections in {\tt HORACE} is  closer to {\tt SANC} than to {\tt PHOTOS}.

Let us stress, that the main purpose of {\tt SANC} is to control dominant,
leading logarithm effects of pairs emission for the sake to supplement
 systematic error evaluation for observables, where pair effects are
 comparable to systematic errors of other effects.
That is why, non leading terms such as $\ln{\frac{\mu}{m_\mu}}\simeq 6$ may be
neglected if they accompany dominant $\ln{\frac{\mu}{m_e}}\simeq 11$
ones. It may be of interest to implement such non-leading terms into {\tt SANC} and/or {\tt PHOTOS}.

\begin{figure}[htp!]
\begin{center}
\begin{minipage}[t]{1.0\textwidth}
\begin{minipage}[h]{0.47\linewidth}
\center{\includegraphics[width=1\linewidth]{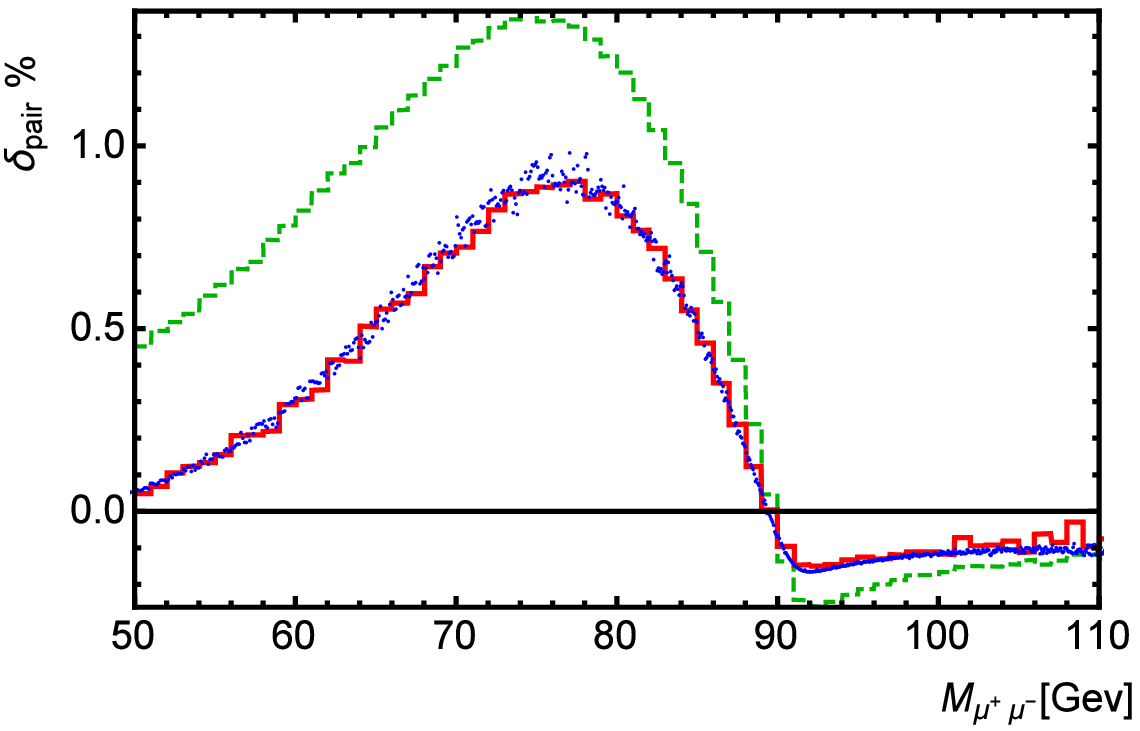}}
\flushleft
a)  Points represent results of simulation by {\tt PYTHIA}, convoluted bin by bin  with our new formula~(\ref{BFactor}).
\\
\end{minipage}
\hfill
\begin{minipage}[h]{0.47\linewidth}
\center{\includegraphics[width=1\linewidth]{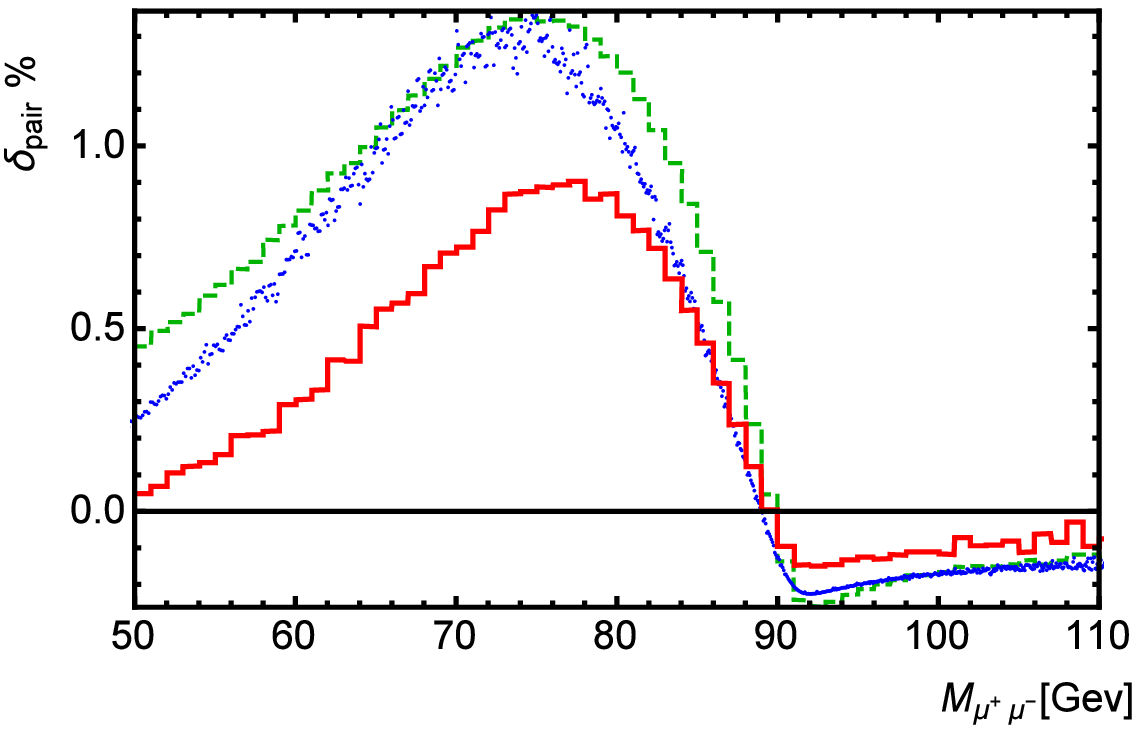}}
\flushleft
b)  Points represent results of simulation by {\tt PYTHIA}, convoluted bin by bin with formula~(\ref{BFactor_Skrzypek}) i.e. as of Ref.~\cite{Jadach:1993wk}.
\\
\end{minipage}
\end{minipage}
\end{center}
\caption[] {\small Comparison of {\tt PHOTOS} and {\tt SANC}  simulations and calculations
 of extra pair emissions, for the process
$p p\rightarrow\nolinebreak Z\nolinebreak\rightarrow\nolinebreak \mu^{+}\mu^{-}\nolinebreak(e^{+}e^{-})$ at $14$~TeV,
with independent semi analytical calculations.
Correction to lepton pair invariant mass spectrum of {\tt PYTHIA} generated sample  is given in \%.
Dashed line represents  {\tt SANC}. Solid line represents data by
{\tt PYTHIA}$\times${\tt PHOTOS}. Numerical results obtained with the help
of formulae~(\ref{BFactor}) or~(\ref{BFactor_Skrzypek}) are superimposed respectively on left and right plot. Our new formula~(\ref{BFactor})
reproduce well  results of {\tt PHOTOS},  but~(\ref{BFactor_Skrzypek}) is closer to results of {\tt SANC}.
\label{fig:mumu_correction}}
\end{figure}

We start semi-analytical tests. Previous researches in this direction can be found in
ref.~\cite{Antropov:2017mwj}. Now we will also use formula~(5) of ref.~\cite{Jadach:1993wk}
(we recall it in Appendix as formula~(\ref{BFactor_Skrzypek})). For its calculation the approximation of factorization
for phase space is used, it is universal and applies to initial state pair emissions as well.
For technical tests of {\tt PHOTOS} and for better understanding of the features of
differences, the semi analytical calculation was repeated, but with exact parametrization of final state emission phase space.
Alternative formula~(\ref{BFactor}) was obtained in Appendix. The numerical tests are
summarized in figs.~\ref{fig:ee_correction} and \ref{fig:mumu_correction}.

\begin{figure}[htp!]
\begin{center}
\begin{minipage}[t]{1.0\textwidth}
\begin{minipage}[h]{0.47\linewidth}
\center{\includegraphics[width=1\linewidth]{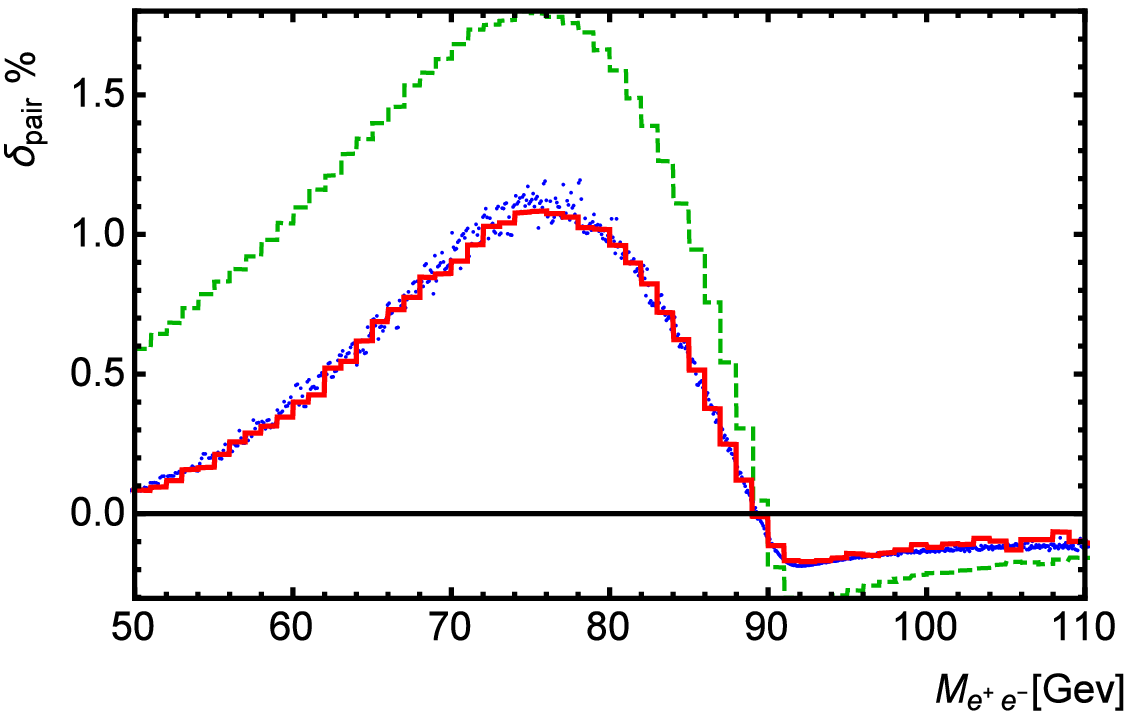}}
\flushleft
a)  Points represent results of simulation by {\tt PYTHIA}, convoluted bin by bin with our new formula~(\ref{BFactor}).
\\
\end{minipage}
\hfill
\begin{minipage}[h]{0.47\linewidth}
\center{\includegraphics[width=1\linewidth]{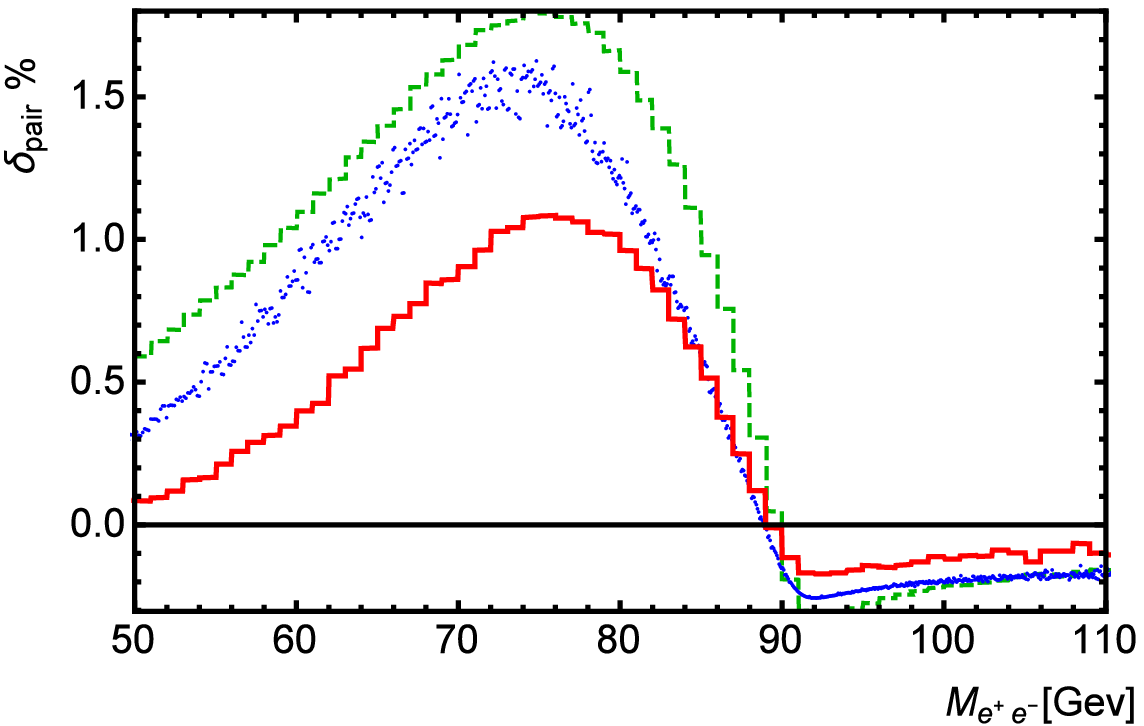}}
\flushleft
b)  Points represent results of simulation by {\tt PYTHIA}, convoluted bin by bin with formula~(\ref{BFactor_Skrzypek}) i.e. as of Ref.~\cite{Jadach:1993wk}.
\\
\end{minipage}
\end{minipage}
\end{center}
\caption[] {\small Comparison of {\tt PHOTOS} and {\tt SANC} simulations and calculations
of extra pair emissions, for the process
$p p\rightarrow\nolinebreak Z\rightarrow\nolinebreak e^{+}e^{-}\nolinebreak(e^{+}e^{-})$ at $14$~TeV,
with independent semi analytical calculations.
Correction to lepton pair invariant mass spectrum of {\tt PYTHIA} generated sample  is given in \%.
Dashed line represents  {\tt SANC}. Solid line represents data by
{\tt PYTHIA}$\times${\tt PHOTOS}. Numerical results obtained with the help
of formulae~(\ref{BFactor}) or~(\ref{BFactor_Skrzypek}) are superimposed respectively on left and right plot. Our new formula~(\ref{BFactor})
reproduce well  results of {\tt PHOTOS},  but~(\ref{BFactor_Skrzypek}) is closer to results of {\tt SANC}.
\label{fig:ee_correction}}
\end{figure}

\begin{itemize}
\item We monitor again, as in Figs.~\ref{fig:diagpair} and \ref{fig:Zee}, the spectrum of invariant mass for the lepton pair, which  is modified by emission of additional pair.
\item For results of {\tt PHOTOS}~\cite{Davidson:2010ew} and for semi-analytical calculation we first generate the sample of events from {\tt PYTHIA}~\cite{Sjostrand:2007gs} with initialization summarized in Fig.~\ref{fig:initialization}.
\item In order to complete results for {\tt PHOTOS}, its algorithm is  applied on events generated by {\tt PYTHIA}.
\item For calculation with formulae~(\ref{BFactor}-\ref{BFactor_Skrzypek}) we move events,
that are generated by {\tt PYTHIA}, to every possible bin of our test distributions with
probabilities obtained from formula~(\ref{BFactor}) or~(\ref{BFactor_Skrzypek}) respectively.
\item Results from {\tt SANC} were obtained earlier and we do not recall
all details necessary for technical
control. They also represent correction for final state emission but spectrum of events
prior emission may differ, because slightly different initialization as of
Fig.~\ref{fig:initialization} was used. Also, instead of formula~(\ref{BFactor_Skrzypek})
equivalent  of formula~(11) as explained in Section \ref{sec:SANC} was used.
Thus some discrepancy is to  be expected.
  \end{itemize}

Analyzing the Fig.~\ref{fig:ee_correction}a, Fig.~\ref{fig:mumu_correction}a we can conclude,
that {\tt PHOTOS} is well in agreement with analytical calculation.
Numerical precision
of agreement is better than $5\%$ of the pair effect. Estimation is limited
by the numerical calculation and CPU time. It can be improved rather easily.
The result is supplemented with Fig.~\ref{fig:narrow_peak} of Appendix,
which is of more technical nature. It includes plots for muon pair emissions.

If instead, results from formula~(\ref{BFactor_Skrzypek}) are used,
see Fig.~\ref{fig:ee_correction}b and Fig.~\ref{fig:mumu_correction}b,
results of {\tt SANC} are much closer than of {\tt PHOTOS} to that variant of
semi-analytical calculation.
Taking all these results together  we can conclude that we understand numerical
difference between  {\tt PHOTOS} and {\tt SANC}.

The main difference between formula~(\ref{BFactor}) and~(\ref{BFactor_Skrzypek}) is
that~(\ref{BFactor}) was obtained by rigorous integration over $4-$body phase space for final
state emissions of matrix element as given in formula~(\ref{sigma}).
For formula~(\ref{BFactor_Skrzypek}) different kinematical conditions
(in fact of initial-state
emissions) were taken into considerations. If energy of the emitted pair is restricted to
soft pair emissions limit, the two calculations coincide, as they should.

One can argue that formula~(\ref{BFactor_Skrzypek}) is less suitable for final state pairs
emissions. This is not necessarily to be the case. For formula~(\ref{BFactor})
  a factorization form of matrix
element is used, but such approximation is not used for phase space. This is potential
source of numerically important mismatches. Even though exact phase space parametrization
offer convenient starting point for future work with matrix element, independent tests with
calculations based on four fermions final state matrix elements are of importance.

\begin{figure}[htp!]
\begin{center}
\begin{minipage}[t]{1.0\textwidth}
\begin{minipage}[h]{0.47\linewidth}
\center{\includegraphics[width=1\linewidth]{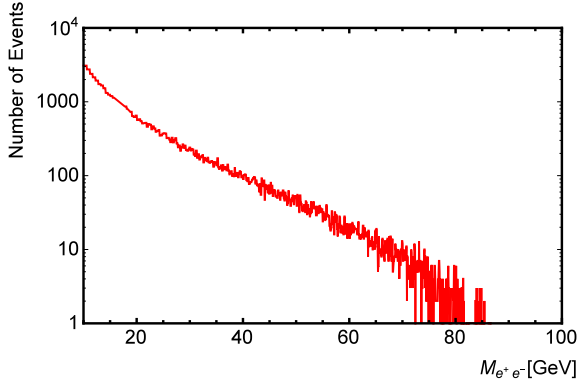}}
\flushleft
a)  $p p\rightarrow Z\nolinebreak\rightarrow\nolinebreak e^{+}e^{-}\nolinebreak(e^{+}e^{-})$;
probability for presence
of  additional pair is  $\simeq 3  \cdot 10^{-3}$.
\\
\end{minipage}
\hfill
\begin{minipage}[h]{0.47\linewidth}
\center{\includegraphics[width=1\linewidth]{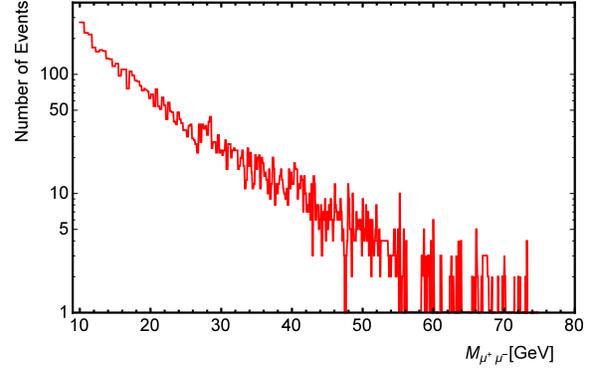}}
\flushleft
b)  $p p\rightarrow Z\rightarrow \mu^{+}\mu^{-}(\mu^{+}\mu^{-})$; probability for presence
of  additional pair is  $\simeq 10^{-4}$.
\\
\end{minipage}
\end{minipage}
\end{center}
\caption[] {Invariant mass distribution in the singlet channel, i.e. of  pair formed from
$l^+$  of emitting pair and $l^-$ of emitted pair generated by {\tt PHOTOS}.  {\tt PYTHIA}
 initialization parameters are presented on Fig.~\ref{fig:initialization}.  Generated samples
(of $\sim 10^8$ events), were dominated by configurations with $M(l^+l^-) \simeq 10$ GeV.

\label{fig:singlet}}
\end{figure}

The {\tt PHOTOS} can be used as well to analyze an effect of singlet channel, which is the case of misidentification in the detector of first lepton as secondary one, when lepton pair emit lepton pair of the same kind.
On Fig.~\ref{fig:singlet}, {\tt PHOTOS} simulations of singlet channel are presented.
Number of events fall down logarithmically with rise of invariant mass of misidentified pair. This perfectly agrees with theory.

\begin{figure}[htp!]
\begin{center}
\begin{minipage}[t]{1.0\textwidth}
\begin{minipage}[h]{0.47\linewidth}
\center{\includegraphics[width=1\linewidth]{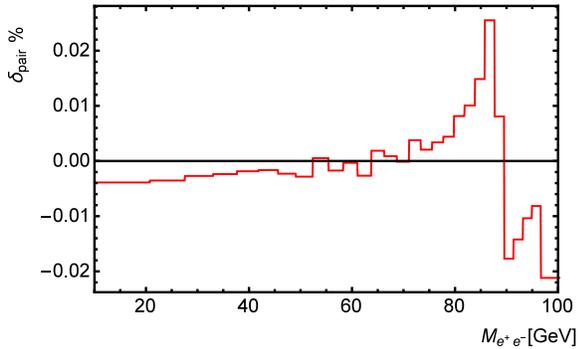}}
\flushleft
a)  $p p\rightarrow Z\nolinebreak\rightarrow\nolinebreak e^{+}e^{-}\nolinebreak(e^{+}e^{-})$.
\\
\end{minipage}
\hfill
\begin{minipage}[h]{0.47\linewidth}
\center{\includegraphics[width=1\linewidth]{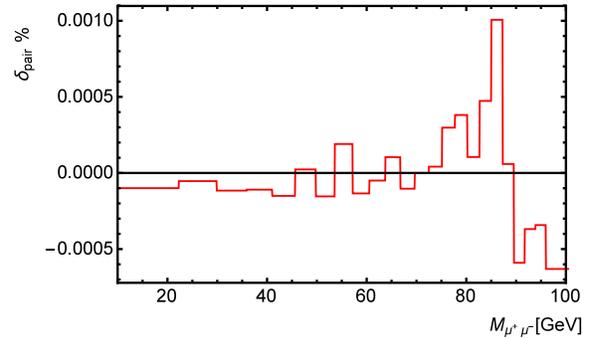}}
\flushleft
b)  $p p\rightarrow Z\rightarrow \mu^{+}\mu^{-}(\mu^{+}\mu^{-})$.
\\
\end{minipage}
\end{minipage}
\end{center}
\caption[] {Pair correction to spectrum of lepton pair invariant mass of {\tt PYTHIA} generated sample is given in \%. Original sample is simulated for pp collisions of $14$~TeV. Solid line represents data by {\tt PYTHIA}$\times${\tt PHOTOS}. Additional lepton pairs are generated under condition that energy of the additional lepton pair in the rest frame of colliding partons is less than $1$~GeV.
\label{fig:Correction1GeV}}
\end{figure}

On Fig.~\ref{fig:Correction1GeV}, soft pair corrections are presented. The cutoff $\Delta=1$~GeV and is applied
for energy of the additional lepton pair in the rest frame of colliding partons. This value for cutoff is chosen both to fulfill the conditions $4 m_f^2\ll\Delta^2\ll M_Z^2$, which correspond to soft pair emissions, and to simulate an effect of the undetected pairs. Depending on the sensitivity of the detector, part of soft lepton pairs remains undetected causing shift in the $p p\rightarrow Z\rightarrow l^{+}l^{-}$ spectrum.

The {\tt KORALW} \cite{Jadach:1998gi} Monte Carlo can be used to generate
$e^+e^- \to 4f$ processes and provide further source of benchmarks
for our studies.  For that purpose it is necessary to run the program for
the Center of Mass
Energy equal to $Z$ boson mass and  $Z$ width set to a very small value,
effectively  to switch off emission of pair from initial state.
Once parameters of pre-sampler adjusted, program was capable of generating
$e^+e^- \to Z \to \mu^+\mu^-\mu^+\mu^-$ or
$e^+e^- \to Z \to \mu^+\mu^-\tau^+\tau^-$ processes over the full phase space.
Once  $m_\tau$ was replaced with electron mass, all necessary
for our testing  options were prepared. For {\tt PHOTOS}
sample leptons  can
originate from emissions or from the pair emitting. In case of  $e^+e^-\mu^+\mu^-$ final state, equal number of
$Z\to e^+e^-$ and $Z\to \mu^+\mu^-$ decays was used. Normalization for the sample size was  fixed
to assure 1M of four-fermion events. Absolute normalization of pair emissions in {\tt PHOTOS} is verified elsewhere, as explained in
Section \ref{sec:PHOTOSpairs},
thanks to tests with analytical formula.

\begin{figure}[htp!]
\begin{minipage}[t]{1.0\textwidth}
\begin{minipage}[h]{0.47\linewidth}
\center{\includegraphics[width=1\linewidth]{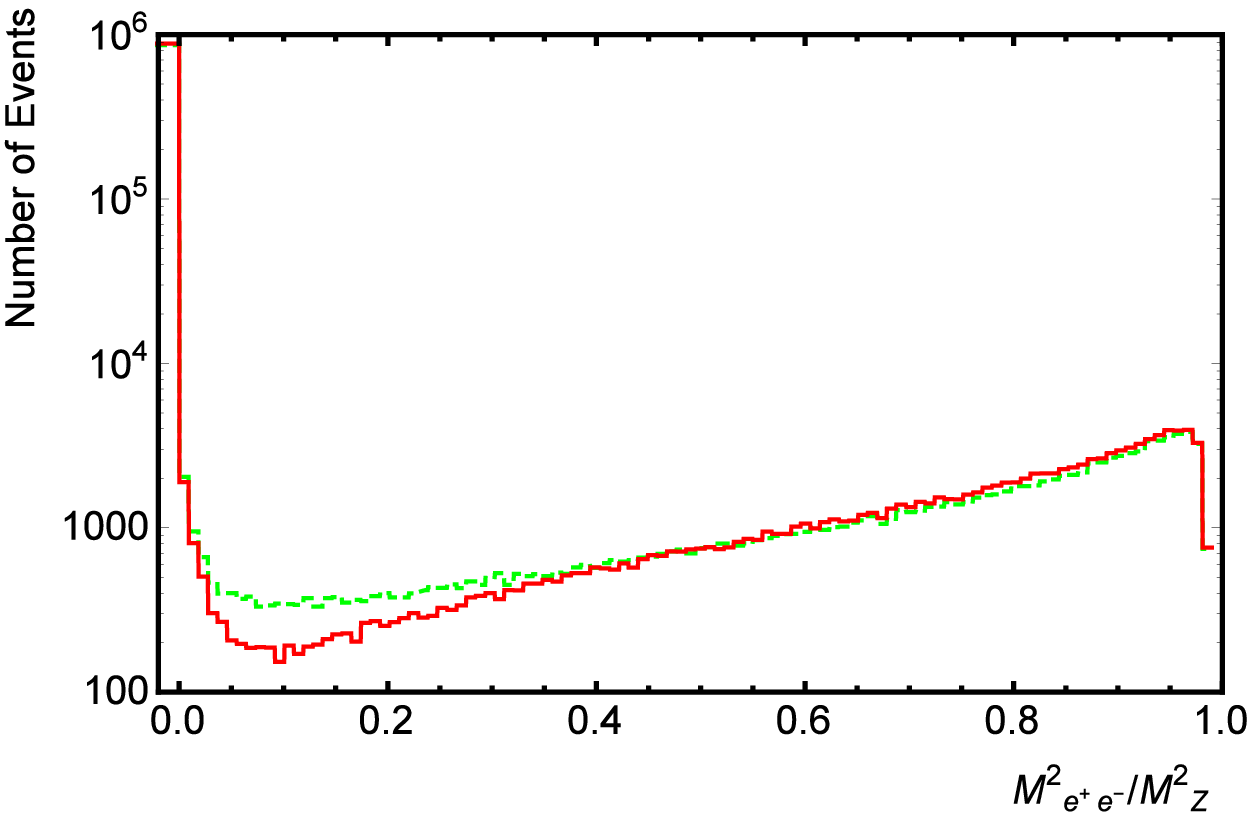}}
\justifying
a)  Normalized to $M_Z^2$ spectrum of electron pair mass squared.
\\
\end{minipage}
\hfill
\begin{minipage}[h]{0.47\linewidth}
\center{\includegraphics[width=1\linewidth]{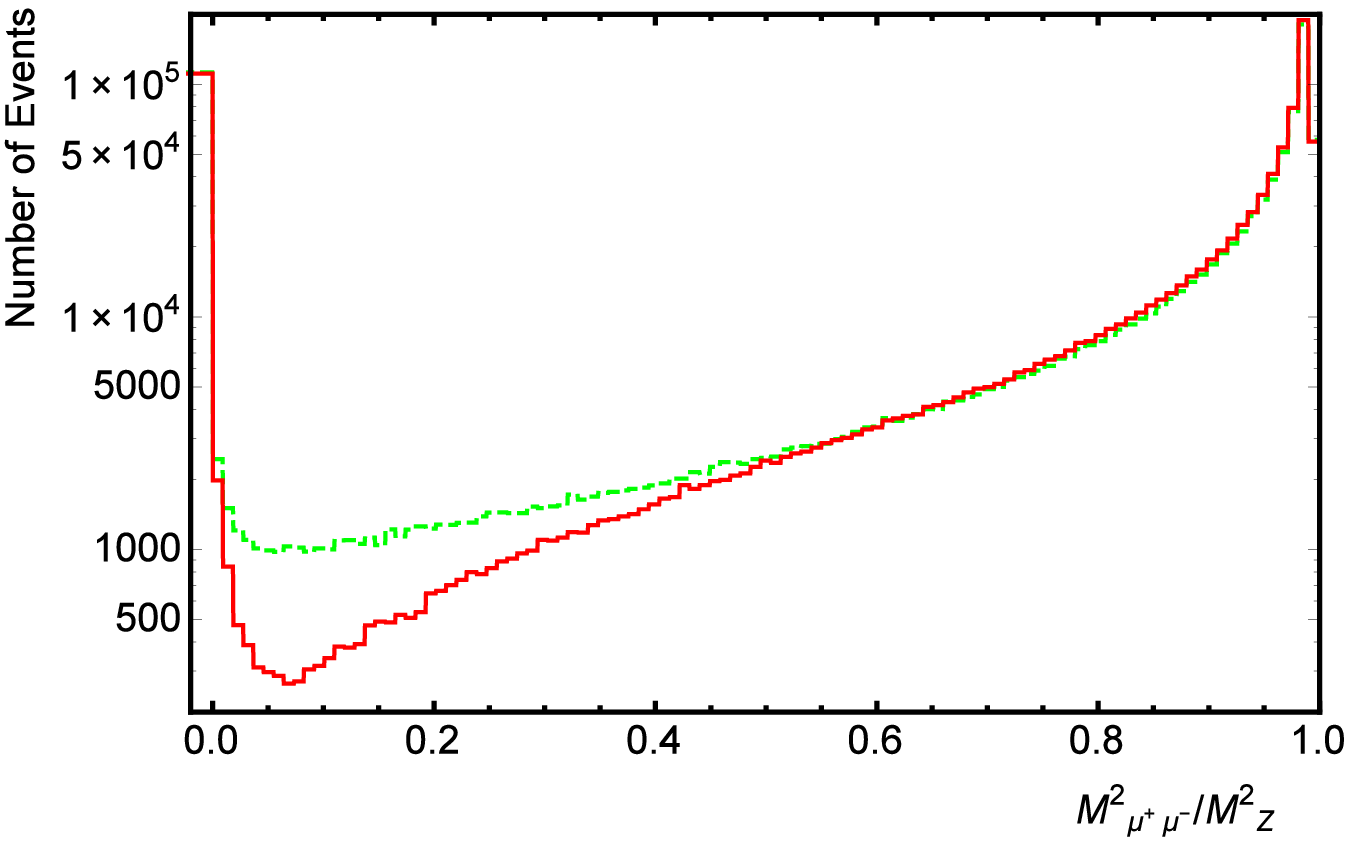}}
\justifying
b)  Normalized to $M_Z^2$ spectrum of  muon pair mass squared.
\\
\end{minipage}
\end{minipage}
\caption[] { \ Lepton pair invariant mass spectra in the channel $Z\to \mu^+\mu^-e^+e^-$.  Results
 generated by {\tt PHOTOS} (solid red line) are obtained from
 samples of equal number of $Z\to\nolinebreak e^+e^-$ and $Z\to \mu^+\mu^-$ decays. They
are compared with results from {\tt KORALW} (dashed green line) where four fermion final state matrix elements are used
as explained in  the text.
{ Agreement of  most populated bins is of importance for test of {\tt PHOTOS}.  }
\label{fig:two-emu}}
\end{figure}

\begin{figure}[htp!]
\begin{minipage}[t]{1.0\textwidth}
\begin{minipage}[h]{0.47\linewidth}
\center{\includegraphics[width=1\linewidth]{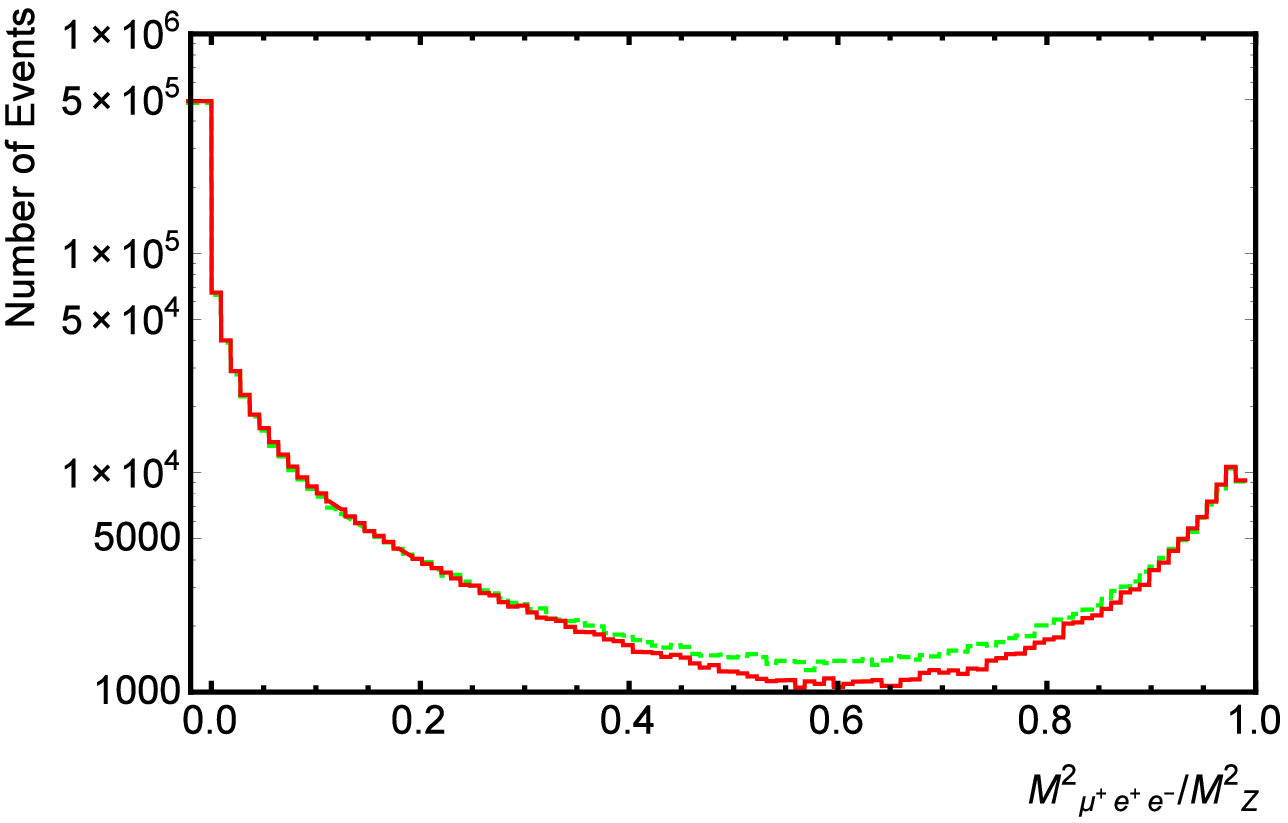}}
\justifying
a)  Normalized to $M_Z^2$ spectrum of  $\mu^+e^+e^-$ mass squared.
\\
\end{minipage}
\hfill
\begin{minipage}[h]{0.47\linewidth}
\center{\includegraphics[width=1\linewidth]{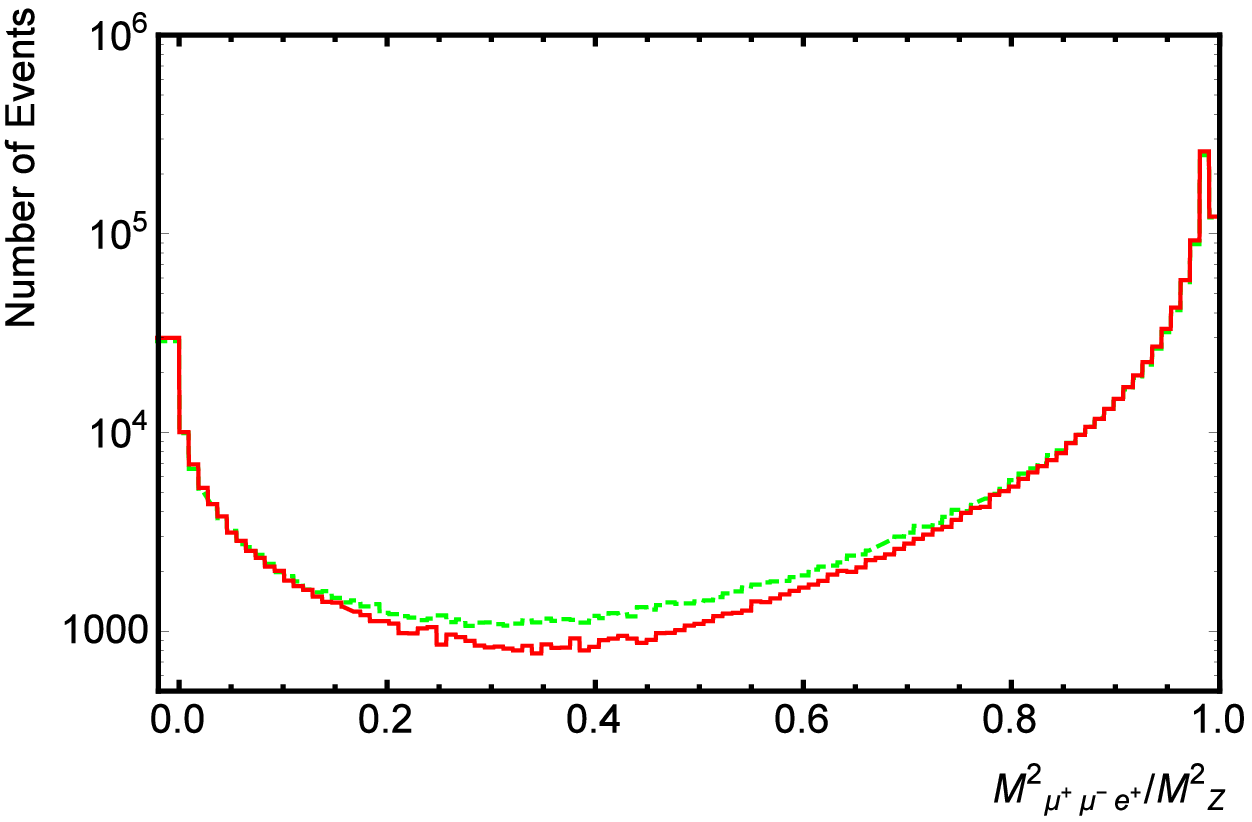}}
\justifying
b)  Normalized to $M_Z^2$ spectrum of  $\mu^+\mu^-e^+$ mass squared.
\\
\end{minipage}
\end{minipage}
\caption[] { \ Invariant mass spectra in the channel $Z\to \mu^+\mu^-e^+e^-$.  Results
 generated by {\tt PHOTOS} (solid red line) are obtained from
 samples of equal number of $Z\to e^+e^-$ and $Z\to \mu^+\mu^-$ decays. They
are compared with results from {\tt KORALW} (dashed green line) where four fermion final state matrix elements are used
as explained in  the text.
 Agreement of  most populated bins is of importance for test of {\tt PHOTOS}.
\label{fig:three-emu}}
\end{figure}

Let us present some numerical results for the samples of 1M events.
In Fig.~\ref{fig:two-emu} we present invariant masses of lepton pairs.
In Fig.~\ref{fig:three-emu} invariant masses for
group of three leptons are shown. This is equivalent, for the dominant contribution,
to test of the angle between emitted pair and one of the original emitters.

\begin{figure}[htp!]
\begin{minipage}[t]{1.0\textwidth}
\begin{minipage}[h]{0.47\linewidth}
\center{\includegraphics[width=1\linewidth]{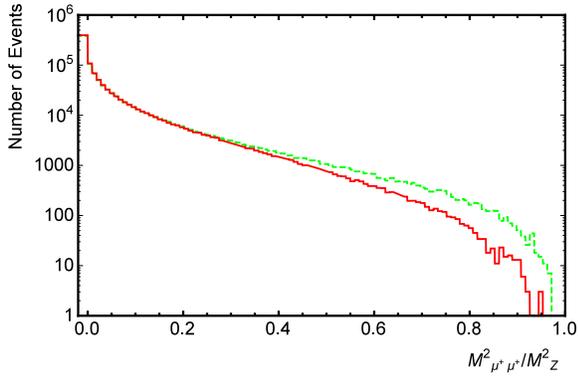}}
\justifying
a)  Normalized to $M_Z^2$ spectrum of  $\mu^+\mu^-$ mass squared.
\\
\end{minipage}
\hfill
\begin{minipage}[h]{0.47\linewidth}
\center{\includegraphics[width=1\linewidth]{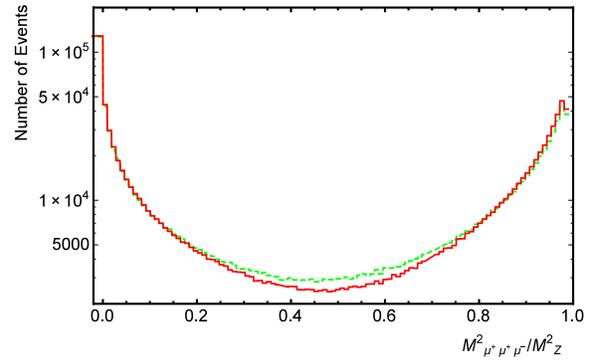}}
\justifying
b)  Normalized to $M_Z^2$ spectrum of  $\mu^+\mu^+\mu^-$ mass squared.
\\
\end{minipage}
\end{minipage}
\caption[] { Invariant mass spectra in the channel $Z\to \mu^+\mu^-\mu^+\mu^-$.  Results
 generated by {\tt PHOTOS} (solid red line) are obtained from
 samples of  $Z\to \mu^+\mu^-$ decays. They
are compared with results from {\tt KORALW} (dashed green line) where four fermion final state matrix elements are used
as explained in  the text.
 Agreement of  most populated bins is of importance for test of {\tt PHOTOS}.
\label{fig:mu}}
\end{figure}

For the muon pair emission in $Z\to \mu^+\mu^-$ we have prepared only one figure
\ref{fig:mu}. Again, reasonable agreement is shown. Further figures, for all invariant masses which can be constructed from  $e^+e^-\mu^+\mu^-$ or
$\mu^+\mu^-\mu^+\mu^-$
are available from the web page \cite{webpage}.

As expected, in some regions of the phase-space, matrix
element based {\tt KORALW} and pair correction kinematics distribution
 generated by {\tt PHOTOS}, vary sizably. This is expected, and of
no significance for establishing precision of  {\tt PHOTOS} as generator
of pair corrections; the corrections which are themselves at the several permille level only,
for the process such as   $Z\to l^+l^-$ decay. For the bins, where bulk
of distribution resides, agreement between {\tt KORALW} and  {\tt PHOTOS}
is at the percent level.

Results of the test are encouraging. Good agreement in the region of phase space of soft
emissions is obtained. For high energy emissions results from {\tt KORALW} seem to indicate for somewhat harder
spectrum than of {\tt PHOTOS}, but not as hard as of {\tt SANC}. This is encouraging observation and clear indication for the future direction of work if higher precision will be needed.

\section{Higher order effects}

Both {\tt SANC} and {\tt PHOTOS} can generate pair effects simultaneously
with emission of photons. Because of rather steep energy spectrum for emitted
pairs, the effect of photonic bremsstrahlung on pair emission
  is not expected to be large. To validate this expectation we have introduced
the following option into  {\tt PHOTOS}; instead of generating in 50 \% of
cases,
pair emission before algorithm for photon emission is involved we have
always generated pairs as the last step. Standard tests with the help of
 {\tt MC-TESTER} demonstrate about 4 \% increase in the number of final
states consisting of configurations with added pair and at least one real
photon of energy above 1 GeV. Shapes of distributions remained not modified
in a noticeable way for the sample of 100 MeV events (see \cite{webpage}).

This provides not only consistency check, but also confirms that {\tt PHOTOS}
can be used with generator such as {\tt KKMC} \cite{Jadach:1999vf} for generation of final state
pair emissions. This, of course, require that intermediate $Z/\gamma^*$ state
is present in the event record. Such intermediate state can be obtained
from the low level generation of {\tt KKMC}. Even if it is not physically
justified to define $Z/\gamma^*$ intermediate state once initial-final
state interference is taken into account, resulting inconsistency is only at
the \% level, at most, of the pair emission effect which itself is at \% level
too. It is thus at the $10^{-4}$ precision level.

\section{Conclusions}
We can conclude that we control bulk of pair effects, down
to 10 \% of their size in the regions of phase space of importance for experimental
conditions, that is for emitted pairs of rather small energies, or collinear.
Rare events featuring hard pairs, could bring larger ambiguities, but are expected also
to be outside of experimental acceptance.
For this region of phase space taken separately, uncertainty is larger, of order of
 even $50\%$, but on the other hand, events of such configurations contribute to the overall Drell-Yan sample at sub-permille level.

The origin of the differences between  {\tt PHOTOS} and {\tt SANC} results used for
the systematic error evaluation is localized and confirmed with semi-analytical calculation.
It is due to approximation resulting from how eq.~(\ref{sigma}) is used in {\tt PHOTOS}
and in {\tt SANC}. Phase space, as used in {\tt PHOTOS} algorithm, is explicit and exact,
enabling for straightforward improvement of matrix element. Note that {\tt PHOTOS}  usage
of approximation in matrix element, but not in phase space, may not be optimal. This is why
solution used in {\tt SANC}, a priori, is not of lower precision than that of {\tt PHOTOS}.
We argue to improve   the precision tag from   0.3\% to  0.1\% for the pair implementation
of the two programs and in applications for observables relevant for heavy boson reconstruction. We provide
 indications for steps necessary to improve beyond 0.1\% precision level.

For the estimation of ambiguities size, the comparisons with {\tt KORALW}, where
complete $2\rightarrow4$ fermion matrix element is available, was instrumental.
It may need to be continued in the future, but as hard pairs contribute to the bulk
of differences, it may not be of urgency for present day experimental effort. This
region of phase-space is expected to remain outside of experimental acceptance.

\vspace*{0.5cm}
{\bf Acknowledgments}

R.S. is grateful for a financial support from ENIGMAS program and to kind hospitality
of the ATLAS group in LAPP.
This work was partially supported by the funds of Polish National Science
Center under decision  UMO-2014/15/B\-/ST2/00049.
Useful discussions with~Maciej~Skrzypek and also his help with {\tt KORALW}
installation are appreciated.
The work is supported in part by the Programme of the French-Polish
Cooperation between  IN2P3 and  COPIN within  the  collaborations  Nos. 10-138 and 11-142.

\bibliographystyle{utphys_spires}
\bibliography{PairEmissionDY}

\appendix
\section{Appendix.}
Let us collect formulae of our calculation used to understand details of analytic calculation of ref.~\cite{Jadach:1993wk}. We have prepared variant of analytic calculation matching solution used in {\tt PHOTOS}. We start from the phase-space parametrization and  integration of matrix element follows.

\subsection{Parametrization of the phase space.}
\begin{eqnarray}\nonumber
\Omega=& &\resizebox{0.8\linewidth}{!}{%
$\displaystyle
\int
\frac{d^3 q_1}{2 (q_1)_0 (2\pi)^3}\cdot\frac{d^3 q_2}{2 (q_2)_0 (2\pi)^3}\cdot\frac{d^3 p}{2 p_0 (2\pi)^3}\cdot\frac{d^3 p'}{2 p'_0 (2\pi)^3} (2\pi)^4\delta^4(R-p-p'-q_1-q_2)=
$}\\\nonumber
&=&\int d^4q d^4Q
\frac{d^3 q_1}{2 (q_1)_0 (2\pi)^3}\cdot\frac{d^3 q_2}{2 (q_2)_0 (2\pi)^3}\cdot\frac{d^3 p}{2 p_0 (2\pi)^3}\cdot\frac{d^3 p'}{2 p'_0 (2\pi)^3} (2\pi)^4\times\\\label{phasespace}
&\times&\delta^4(R-p-p'-q_1-q_2)\delta^4(q-q_1-q_2)\delta^4(Q-p-p')
\end{eqnarray}

\begin{eqnarray}
\int \frac{d^3 q_1}{2 (q_1)^0} \frac{d^3 q_2}{2 (q_2)^0} \delta^4(q-q_1-q_2)=\int \frac{|\overline{q_1}| d\cos\theta_{q_1} d\phi_{q_1}} {4\sqrt{q^2}},
\end{eqnarray}
where $\theta_{q_1}, \phi_{q_1}$ are direction of $q_1$ in the rest frame of $q$, $|\overline{q_1}|=|\overline{q_2}|=\sqrt{\frac{q^2}4-\mu^2}$.

\begin{eqnarray}
\int \frac{d^3 p}{2 (p)^0} \frac{d^3 p'}{2 (p')^0} \delta^4(Q-p-p')=\int \frac{|\overline{p}| d\cos\theta_{p} d\phi_{p}} {4\sqrt{p^2}},
\end{eqnarray}
where $\theta_{p}, \phi_{p}$ are direction of $p$ in the rest frame of $Q$, $|\overline{p}|=|\overline{p'}|=\sqrt{\frac{Q^2}4-m^2}$.

\begin{eqnarray}
& &\int d^4q d^4Q \delta^4(R-Q-q)=\int (d\cos\theta_qd\phi_q) d M^2_Q d M^2_q \frac{\sqrt{\lambda}}{8s}
\end{eqnarray}
where $\theta_{q}, \phi_{q}$ are direction of $q$ in the rest frame of $R$.

\begin{eqnarray}\label{Phase_space}
\resizebox{0.9\linewidth}{!}{%
$\displaystyle
\Omega=\frac1{(2\pi)^8}\int d M_q^2 d M_Q^2 d cos\theta_{q_1} d\phi_{q_1} d cos\theta_{p} d\phi_{p} d cos\theta_{q} d\phi_{q} \frac18\sqrt{1-\frac{4\mu^2}{q^2}}\frac18\sqrt{1-\frac{4m^2}{Q^2}}\frac{\sqrt{\lambda(s,M_Q^2,M_q^2)}}{8s}.
$}
\end{eqnarray}

We choose that:
\begin{enumerate}
\item $\theta_{p}, \phi_{p}$ define orientation of $p$ (in the rest frame of $Q$) with respect to $z$ axis along direction of q (as seen in this frame);
\item $\theta_{q_1}, \phi_{q_1}$ define orientation of $q_1$ (in the rest frame of $q$) with respect to $z$ axis along boost from this frame to the rest frame of $Q$;
\item $\theta_{q}, \phi_{q}$ define orientation of $p$ with respect to laboratory directions (in the rest frame of $R$).
\end{enumerate}

\subsection{Preparation of the Matrix Element.}

Let us now turn our attention to matrix element.
Factorized term obtained from pair emission matrix element and used in ref.~\cite{Jadach:1993wk} formula (1) as integrand reads:
\begin{eqnarray}\label{MatrixElement}
&F(p,p',q,q_1,q_2,a)=(\frac{\alpha}{\pi})^2\frac1{\pi^2}\left( \frac{2p-a q}{a q^2 -2 p q} - \frac{2p'-a q}{a q^2 -2 p' q}\right)_\mu \left( \frac{2p-a q}{a q^2 -2 p q} - \frac{2p'-a q}{a q^2 -2 p' q}\right)_\nu \frac{4 q_1^\mu q_2^\nu - q^2 g^{\mu\nu}}{2 q^4}.
\end{eqnarray}
Note that it includes  factor $\frac{1}{(2\pi)^6}$ of the phase-space integration  volume.
We need to recall that at the end of calculation.

Now we can express all four vectors necessary for formula~(\ref{MatrixElement})
with the help of previously specified angles.
Four vectors $p,p',q,q_1,q_2$ in the rest frame of $Q$ read:
\begin{eqnarray}\nonumber
p&=&(E_p,p\cos \phi_p \sin \theta_p,p\sin \phi_p \sin \theta_p,p \cos \theta_p),\\\nonumber
p'&=&(E_p,-p\cos \phi_p \sin \theta_p,-p\sin \phi_p \sin \theta_p,-p \cos \theta_p),\\\label{Four_Vectors}
q&=&(E_q,0,0,q),
\end{eqnarray}
where
\begin{eqnarray}\nonumber
E_p&=&\frac12M_Q,\\\nonumber
p&=&\sqrt{\frac{M_Q^2}4-m^2},\\\nonumber
E_q&=&\frac{s-M_Q^2-M_q^2}{2M_Q},\\\label{energies}
q&=&\frac{\sqrt{(s-M_Q^2-M_q^2)^2-4M_Q^2M_q^2}}{2M_Q}.
\end{eqnarray}
To  obtain expressions for  $E_q$ and $q$ formulae  for $p$ and $p'$ and  $s=(p+p'+q)^2$ are needed.

We first define $q_1$ and $q_2$ in the the rest frame of $q$:
\begin{eqnarray}\nonumber
q_1&=&(\frac{M_q}2,v\cos \phi_{q_1} \sin \theta_{q_1},v\sin \phi_{q_1} \sin \theta_{q_1},v \cos \theta_{q_1}),\\\nonumber
q_2&=&(\frac{M_q}2,-v\cos \phi_{q_1} \sin \theta_{q_1},-v\sin \phi_{q_1} \sin \theta_{q_1},-v \cos \theta_{q_1}),
\end{eqnarray}
where
\begin{eqnarray}
v&=&\sqrt{\frac{M_q^2}4-\mu^2}.
\end{eqnarray}

\subsection{Integration of matrix element.}

We have to calculate
\begin{eqnarray}\label{sigma}
\sigma=\int d \Omega F |M_B|^2,
\end{eqnarray}
where $F$ if given by formula~(\ref{MatrixElement}) and $d \Omega$ by~(\ref{Phase_space}). $|M_B|^2$ is not important as we will see.

Question is how to do it in most convenient way without loosing symmetry properties of~(\ref{MatrixElement}).

Observation:
\begin{enumerate}
\item $F$ depends on all variables except $\theta_q,\phi_q$;
\item $|M_B|^2$ depends only on $\theta_q,\phi_q$;
\item $\theta_{q_1},\phi_{q_1}$ are present only in $\frac{4 q_1^\mu q_2^\nu - q^2 g^{\mu\nu}}{2 q^4}$.
\end{enumerate}
It is convenient to integrate $\frac{4 q_1^\mu q_2^\nu - q^2 g^{\mu\nu}}{2 q^4}$ over $\theta_{q_1},\phi_{q_1}$ in the rest frame of $q$. Because of Lorentz invariance we have
\begin{eqnarray}
\int d\theta_q d\phi_q d \frac{4 q_1^\mu q_2^\nu - q^2 g^{\mu\nu}}{2 q^4} = X g^{\mu\nu}+Y q^\mu q^\nu.
\end{eqnarray}
Thus

\begin{eqnarray}\nonumber
& &\int d\theta_q d\phi_q d \frac{4 q_1^\mu q_2^\nu - q^2 g^{\mu\nu}}{2 q^4} =\\\nonumber
&=&\frac{16\pi}{2M_q^4}
\resizebox{0.4\linewidth}{!}{%
$\displaystyle
\left(
  \begin{array}{cccc}
    \frac{M_q^2}4 & 0 & 0 & 0 \\
    0 & -\frac13\left(\frac{M_q^2}{4}-\mu^2\right) & 0 & 0 \\
    0 & 0 & -\frac13\left(\frac{M_q^2}{4}-\mu^2\right) & 0 \\
    0 & 0 & 0 & -\frac13\left(\frac{M_q^2}{4}-\mu^2\right) \\
  \end{array}
\right)
$}
-
\frac{4\pi M_q^2}{2M_q^4}
\left(
  \begin{array}{cccc}
    1 & 0 & 0 & 0 \\
    0 & -1 & 0 & 0 \\
    0 & 0 & -1 & 0 \\
    0 & 0 & 0 & -1 \\
  \end{array}
\right)=\\\nonumber
&=&\frac1{M_q^2}
\resizebox{0.4\linewidth}{!}{%
$\displaystyle
\left(
  \begin{array}{cccc}
    0 & 0 & 0 & 0 \\
    0 & \frac{4\pi}3\left(1+\frac{2\mu^2}{M_q^2}\right) & 0 & 0 \\
    0 & 0 & \frac{4\pi}3\left(1+\frac{2\mu^2}{M_q^2}\right) & 0 \\
    0 & 0 & 0 & \frac{4\pi}3\left(1+\frac{2\mu^2}{M_q^2}\right) \\
  \end{array}
\right)
$}
=\\\nonumber
&=&\resizebox{0.8\linewidth}{!}{%
$\displaystyle
-\frac1{M_q^2}\cdot\frac{4\pi}3\left(1+\frac{2\mu^2}{M_q^2}\right)
\left(
  \begin{array}{cccc}
    1 & 0 & 0 & 0 \\
    0 & -1 & 0 & 0 \\
    0 & 0 & -1 & 0 \\
    0 & 0 & 0 & -1 \\
  \end{array}
\right)
+\frac1{M_q^2}\cdot\frac{4\pi}3\left(1+\frac{2\mu^2}{M_q^2}\right)
\left(
  \begin{array}{cccc}
    1 & 0 & 0 & 0 \\
    0 & 0 & 0 & 0 \\
    0 & 0 & 0 & 0 \\
    0 & 0 & 0 & 0 \\
  \end{array}
\right)=
$}\\
&=&-\frac1{M_q^2}\cdot\frac{4\pi}3\left(1+\frac{2\mu^2}{M_q^2}\right)g^{\mu\nu}+
\frac1{M_q^2}\cdot\frac{4\pi}3\left(1+\frac{2\mu^2}{M_q^2}\right)\frac{q^{\mu}q^{\nu}}{M_q^2}. \label{eq:virtPH}
\end{eqnarray}
It is easy to verify, that
\begin{eqnarray}
\left( \frac{2p-a q}{a q^2 -2 p q} - \frac{2p'-a q}{a q^2 -2 p' q}\right)_{\mu}
\left( \frac{2p-a q}{a q^2 -2 p q} - \frac{2p'-a q}{a q^2 -2 p' q}\right)_{\nu}
q^{\mu}q^{\nu}
\end{eqnarray}
equals zero, and second part of (\ref{eq:virtPH}) does not contribute.
This is a consequence of property resulting from Ward identity of QED~\cite{Ward:1950xp}.

Products of four-vectors can be expressed with the help of invariants
and masses used in phase-space parametrization
\begin{eqnarray}\nonumber
p\cdot p'&=&\frac{M_Q^2}2-m^2;\\\nonumber
p\cdot q&=&\frac{s-M_Q^2-M_q^2}{4}-\sqrt{\frac{M_Q^2}4-m^2} \frac{\lambda^{\frac12}(s,M_Q^2,M_q^2)}{2M_Q} \cos \theta_p;\\
p'\cdot q&=&\frac{s-M_Q^2-M_q^2}{4}+\sqrt{\frac{M_Q^2}4-m^2} \frac{\lambda^{\frac12}(s,M_Q^2,M_q^2)}{2M_Q} \cos \theta_p. \label{simple_bracket}
\end{eqnarray}
In case of $a=0$ calculation is particularly simple:
\begin{eqnarray}\nonumber
& &\left( \frac{2p-a q}{a q^2 -2 p q} - \frac{2p'-a q}{a q^2 -2 p'q}\right)^2=\\\nonumber
&=&\frac{4 m^2}{\left(\frac{s-M_Q^2-M_q^2}{2}-\sqrt{\frac{M_Q^2}4-m^2} \frac{\lambda^{\frac12}(s,M_Q^2,M_q^2)}{M_Q} \cos \theta_p\right)^2}+\\\nonumber
&+&\frac{4 m^2}{\left(\frac{s-M_Q^2-M_q^2}{2}+\sqrt{\frac{M_Q^2}4-m^2} \frac{\lambda^{\frac12}(s,M_Q^2,M_q^2)}{M_Q} \cos \theta_p\right)^2}-\\\label{square_braket2}
&-&2\frac{2M_Q^2-4m^2}{\frac{(s-M_Q^2-M_q^2)^2}{4}-\left(\frac{M_Q^2}4-m^2\right)
\frac{\lambda(s,M_Q^2,M_q^2)}{M_Q^2}\cos^2\theta_p}.
\end{eqnarray}

In general case thanks to  ~(\ref{Four_Vectors}) we obtain

{\small
\begin{eqnarray}\nonumber
& &
\left( \frac{2p-a q}{a q^2 -2 p q} - \frac{2p'-a q}{a q^2 -2 p'q}\right)^2=
\Bigg{(}\frac{4 p^{\mu}p_{\mu} + a^2 q^{\mu}q_{\mu}-4 a p_{\mu}q^{\mu}}{(a q^{\mu}q_{\mu} -2 E_p E_q+2pq\cos\theta_p)^2}+
\frac{4 p^{\mu}p_{\mu} + a^2 q^{\mu}q_{\mu}-4 a p'_{\mu}q^{\mu}}{(a q^{\mu}q_{\mu} -2 E_p E_q-2pq\cos\theta_p)^2}
\\\nonumber
&&\;\;\; -2\frac{4 p^{\mu}p'_{\mu} -2 a q_{\mu}(p+p')^{\mu} + a^2 q^{\mu}q_{\mu}}{(a q^{\mu}q_{\mu} -2 E_p E_q+2pq\cos\theta_p)(a q^{\mu}q_{\mu} -2 E_p E_q - 2pq\cos\theta_p)}\Bigg{)}\\\nonumber
&&=\Bigg{(}\frac{4 m^2 + a M_q^2-4 a E_p E_q +4 a pq\cos\theta_p}{(a M_q^2 -2 E_p E_q+2pq\cos\theta_p)^2}\\\label{square_braket}
&&\;\;\;+\frac{4 m^2 + a M_q^2-4 a E_p E_q -4 a pq\cos\theta_p}{(a M_q^2 -2 E_p E_q-2pq\cos\theta_p)^2}-
2\frac{4 (m^2 +2p^2) -4 a E_q E_p + a^2 M_q^2}{(a M_q^2 -2 E_p E_q)^2-4p^2q^2\cos^2\theta_p}\Bigg{)}.
\end{eqnarray}
}

In order to integrate expression~(\ref{square_braket}) over $\cos\theta_p$ we
separate it into three parts,  corresponding to distinct polynomials in
$\cos\theta_p$. Integrals  read:

\begin{eqnarray}\nonumber
C_1&=&\int\limits_{1}^{-1} d \cos\theta_p
\left(\frac{4m^2 + a M^2_q-4a E_pE_q}
{(a M^2_q -2 E_p E_q+2pq\cos\theta_p)^2}+\frac{4m^2 + a M^2_q-4a E_pE_q}
{(a M^2_q -2 E_p E_q-2pq\cos\theta_p)^2}\right);\\\nonumber
C_2&=&\int\limits_{1}^{-1} d \cos\theta_p
\left(\frac{4 a pq\cos\theta_p}
{(a M^2_q -2 E_p E_q+2pq\cos\theta_p)^2}-
\frac{4 apq\cos\theta_p}
{(a M^2_q -2 E_p E_q-2pq\cos\theta_p)^2}\right);\\
C_3&=&\int\limits_{1}^{-1} d \cos\theta_p
\frac{4 (m^2 +2p^2) -4 a E_q E_p + a^2 M_q^2}
{(a M_q^2 -2 E_p E_q)^2- 4p^2q^2\cos^2\theta_p}.
\end{eqnarray}

Let us now return to our main eq.~(\ref{sigma}). We get
\begin{eqnarray}\nonumber
\sigma&=&\resizebox{0.9\linewidth}{!}{%
$\displaystyle
\frac1{(2\pi)^8}\frac1{\pi^2}\int|M_B|^2 d M_q^2 d M_Q^2 d cos\theta_{p} d\phi_{p} d cos\theta_{q} d\phi_{q} \frac18\sqrt{1-\frac{4\mu^2}{q^2}}\frac18\sqrt{1-\frac{4m^2}{Q^2}}\frac{\sqrt{\lambda(s,M_Q^2,M_q^2)}}{8s}\times
$}\\\label{sigma_one_integration}
&\times&\resizebox{0.8\linewidth}{!}{%
$\displaystyle
(\frac{\alpha}{\pi})^2
\left( \frac{2p-a q}{a q^2 -2 p q} - \frac{2p'-a q}{a q^2 -2 p' q}\right)_\mu
\left( \frac{2p-a q}{a q^2 -2 p q} - \frac{2p'-a q}{a q^2 -2 p' q}\right)^\mu
\frac1{M_q^2}\cdot\frac{(-4\pi)}3\left(1+\frac{2\mu^2}{M_q^2}\right)
$}
\end{eqnarray}
or after re-ordering of terms
\begin{eqnarray}\nonumber
\sigma&=&\resizebox{0.8\linewidth}{!}{%
$\displaystyle
-\frac1{3\cdot2^{15}\pi^9 s}(\frac{\alpha}{\pi})^2\int\left[|M_B|^2 d cos\theta_{q} d\phi_{q}\right] d M_Q^2 \frac{d M_q^2}{M_q^2}  d cos\theta_{p} d\phi_{p}  \sqrt{1-\frac{4\mu^2}{M_q^2}}\left(1+\frac{2\mu^2}{M_q^2}\right)\sqrt{1-\frac{4m^2}{M_Q^2}}\times
$}\\\label{sigma_one_integration2}
&\times&\resizebox{0.8\linewidth}{!}{%
$\displaystyle\lambda^{\frac12}(s,M_Q^2,M_q^2)
\left( \frac{2p-a q}{a q^2 -2 p q} - \frac{2p'-a q}{a q^2 -2 p' q}\right)_\mu
\left( \frac{2p-a q}{a q^2 -2 p q} - \frac{2p'-a q}{a q^2 -2 p' q}\right)^\mu.
$}
\end{eqnarray}

We simplify  integral~(\ref{sigma_one_integration2}) with the
help of~(\ref{square_braket2}).
Expression~(\ref{simple_bracket}) or(\ref{square_braket})  does not depend
on $\phi_p$,  integration  over $\phi_p$ is trivial and gives an overall factor $2\pi$.
One also notice that integrals over $\cos\theta_p$ of first and second part
 of~(\ref{square_braket2}) are equal. We obtain

\begin{eqnarray}\nonumber
\sigma&=&\resizebox{0.8\linewidth}{!}{%
$\displaystyle
-\frac1{3\cdot2^{15}\pi^9 s}(\frac{\alpha}{\pi})^2\int\left[|M_B|^2 d cos\theta_{q} d\phi_{q}\right] d M_Q^2 \frac{d M_q^2}{M_q^2}  2 \pi   \sqrt{1-\frac{4\mu^2}{M_q^2}}\left(1+\frac{2\mu^2}{M_q^2}\right)\sqrt{1-\frac{4m^2}{M_Q^2}}\times
$}\\\nonumber
&\times&\lambda^{\frac12}(s,M_Q^2,M_q^2)
\int\limits_{1}^{-1}d cos\theta_{p}
\Bigg{[}\frac{8 m^2}{\left(\frac{s-M_Q^2-M_q^2}{2}-\sqrt{\frac{M_Q^2}4-m^2} \frac{\lambda^{\frac12}(s,M_Q^2,M_q^2)}{M_Q} \cos \theta_p\right)^2}-\\\label{sigma_one_integration3}
&-&2\frac{2M_Q^2-4m^2}{\frac{(s-M_Q^2-M_q^2)^2}{4}-\left(\frac{M_Q^2}4-m^2\right)
\frac{\lambda(s,M_Q^2,M_q^2)}{M_Q^2}\cos^2\theta_p}\Bigg{]}.
\end{eqnarray}

Now we need to integrate over $\cos\theta_p$. The following formulas are helpful
\begin{eqnarray}\nonumber
\int\limits_{-1}^{1}\frac{d x}{(A-B x)^2}=
\frac{2}{A^2-B^2}
\end{eqnarray}
and
\begin{eqnarray}\nonumber
\int\limits_{-1}^{1}\frac{dx}{A^2-B^2 x^2}=-\frac1{A B}\ln \frac{A-B}{A+B}.
\end{eqnarray}

With help of these, we get:

\begin{eqnarray}\nonumber
\sigma&=&\resizebox{0.8\linewidth}{!}{%
$\displaystyle
-\frac1{3\cdot2^{15}\pi^9 s}(\frac{\alpha}{\pi})^2\int\left[|M_B|^2 d cos\theta_{q} d\phi_{q}\right] d M_Q^2 \frac{d M_q^2}{M_q^2}  2 \pi   \sqrt{1-\frac{4\mu^2}{M_q^2}}\left(1+\frac{2\mu^2}{M_q^2}\right)\sqrt{1-\frac{4m^2}{M_Q^2}}\times
$}\\\nonumber
&\times&\lambda^{\frac12}(s,M_Q^2,M_q^2)
\Bigg{[}\frac{16 m^2}{\frac{(s-M_Q^2-M_q^2)^2}{4}-\left(\frac{M_Q^2}4-m^2\right) \frac{\lambda(s,M_Q^2,M_q^2)}{M_Q^2} }+\\\label{sigma_one_integration4}
&+&2\frac{2M_Q^2-4m^2}{\frac{s-M_Q^2-M_q^2}{2}\sqrt{\frac{M_Q^2}4-m^2}
\frac{\lambda^{\frac12}(s,M_Q^2,M_q^2)}{M_Q}}
\ln\frac{\frac{s-M_Q^2-M_q^2}{2}-\sqrt{\frac{M_Q^2}4-m^2}
\frac{\lambda^{\frac12}(s,M_Q^2,M_q^2)}{M_Q}}{\frac{s-M_Q^2-M_q^2}{2}+\sqrt{\frac{M_Q^2}4-m^2}
\frac{\lambda^{\frac12}(s,M_Q^2,M_q^2)}{M_Q}}
\Bigg{]}.
\end{eqnarray}

Some ordering of terms gives
\begin{eqnarray}\nonumber
\sigma&=&-\frac1{3\cdot2^{10}\pi^8 s}(\frac{\alpha}{\pi})^2\int\left[|M_B|^2 d cos\theta_{q} d\phi_{q}\right] d M_Q^2 \frac{d M_q^2}{M_q^2}      \sqrt{1-\frac{4\mu^2}{M_q^2}}\left(1+\frac{2\mu^2}{M_q^2}\right)\sqrt{1-\frac{4m^2}{M_Q^2}}\times\\\nonumber
&\times&\lambda^{\frac12}(s,M_Q^2,M_q^2)
\Bigg{[}\frac{m^2}{M_q^2 M_Q^2+\frac{m^2}{M_Q^2}\lambda(s,M_Q^2,M_q^2)}+\\\label{sigma_one_integration5}
&+&\resizebox{0.85\linewidth}{!}{%
$\displaystyle
\frac{M_Q^2-2m^2}{(s-M_Q^2-M_q^2)\sqrt{1-\frac{4m^2}{M_Q^2}}
\lambda^{\frac12}(s,M_Q^2,M_q^2)}
\ln\frac{s-M_Q^2-M_q^2-\sqrt{1-\frac{4m^2}{M_Q^2}}
\lambda^{\frac12}(s,M_Q^2,M_q^2)}
{s-M_Q^2-M_q^2+\sqrt{1-\frac{4m^2}{M_Q^2}}
\lambda^{\frac12}(s,M_Q^2,M_q^2)}
\Bigg{]}
$},
\end{eqnarray}

or with explicit expression of Born separated (two body phase space is taken from formula~(36) of ref.~\cite{Was:1994kg}):
\begin{eqnarray}\nonumber
\sigma&=&\frac1{(2\pi)^6}\int\left[\frac1{(2\pi)^2}\cdot\frac{\lambda^{\frac12}(1,\frac{m^2}{s},\frac{m^2}{s})}8|M_B|^2 d cos\theta_{q} d\phi_{q}\right]\times
\;\;\lambda^{-\frac12}(1,\frac{m^2}{s},\frac{m^2}{s})\\\nonumber
&\times&\frac{(-2)}{3s}(\frac{\alpha}{\pi})^2 \int d M_Q^2\frac{d M_q^2}{M_q^2}\sqrt{1-\frac{4 \mu^2}{M_q^2}} \left(1+\frac{2 \mu^2}{M_q^2}\right)\times\Bigg{[}\;\frac{m^2  \sqrt{1-\frac{4 m^2}{M_Q^2}} \lambda^{\frac12}(s,M_Q^2,M_q^2)}{M_q^2 M_Q^2+\frac{m^2}{M_Q^2}\lambda(s,M_Q^2,M_q^2)}\\\label{sigma_one_integration6}
&+&\frac{M_Q^2-2 m^2 }{s-M_q^2-M_Q^2}
\ln \frac{s-M_q^2-M_Q^2-\sqrt{1-\frac{4 m^2}{M_Q^2}}\lambda^{\frac12}(s,M_Q^2,M_q^2)}
{s-M_q^2-M_Q^2+\sqrt{1-\frac{4 m^2}{M_Q^2}}\lambda^{\frac12}(s,M_Q^2,M_q^2)}\;\Bigg{]}.
\end{eqnarray}
\subsection{Result.}

From (\ref{sigma_one_integration6}) we obtain analog of formula~(5) of ref.~\cite{Jadach:1993wk}:

\begin{eqnarray}\nonumber
\widetilde{B}_f&=&-\frac2{3s}(\frac{\alpha}{\pi})^2 \int d M_Q^2\frac{d M_q^2}{M_q^2}\sqrt{1-\frac{4 \mu^2}{M_q^2}} \left(1+\frac{2 \mu^2}{M_q^2}\right) \Bigg{(}\frac{m^2  \sqrt{1-\frac{4 m^2}{M_Q^2}} \lambda^{\frac12}(s,M_Q^2,M_q^2)}{M_q^2 M_Q^2+\frac{m^2}{M_Q^2}\lambda(s,M_Q^2,M_q^2)}+\\\label{BFactor}
&+&\frac{M_Q^2-2 m^2 }{s-M_q^2-M_Q^2}
\ln \frac{s-M_q^2-M_Q^2-\sqrt{1-\frac{4 m^2}{M_Q^2}}\lambda^{\frac12}(s,M_Q^2,M_q^2)}
{s-M_q^2-M_Q^2+\sqrt{1-\frac{4 m^2}{M_Q^2}}\lambda^{\frac12}(s,M_Q^2,M_q^2)}\Bigg{)}
\end{eqnarray}
Note that  the factor $\frac1{(2\pi)^6}$ had to be dropped out to avoid double counting. This factor of phase space
 parametrization was already incorporated into the formula (\ref{MatrixElement}).

In order to make comparison with older calculations, we recall formula~(5)
of ref.~\cite{Jadach:1993wk}; case of $a=0$, which is exact for the
emission of extra lepton pair from initial state.

\begin{eqnarray}\nonumber
\widetilde{B_f}&=&
-\frac2{3s}
(\frac{\alpha}{\pi})^2\int d M^2_Q \frac{d M_q^2}{M_q^2}
\sqrt{1-\frac{4\mu^2}{M_q^2}}\left(1+\frac{2\mu^2}{M_q^2}\right)
\Bigg{(}
\frac{m^2\lambda^{\frac12}(s,M_Q^2,M_q^2)}{ M_q^2 s+\frac{m^2}s \lambda(s,M_Q^2,M_q^2)}+\\\label{BFactor_Skrzypek}
&+&\frac{s-2 m^2}{\sqrt{1-\frac{4m^2}{s}}(s+M_q^2-M_Q^2)}
\ln \frac{s+M_q^2-M_Q^2-\sqrt{1-\frac{4m^2}{s}}\lambda^{\frac12}(s,M_Q^2,M_q^2)}
{s+M_q^2-M_Q^2+\sqrt{1-\frac{4m^2}{s}}\lambda^{\frac12}(s,M_Q^2,M_q^2)}
\Bigg{)}.
\end{eqnarray}

We have now collected all formulae necessary for numerical results.



\begin{figure}[htp!]
\begin{center}
\begin{minipage}[t]{1.0\textwidth}
\begin{minipage}[h]{0.47\linewidth}
\center{\includegraphics[width=1\linewidth]{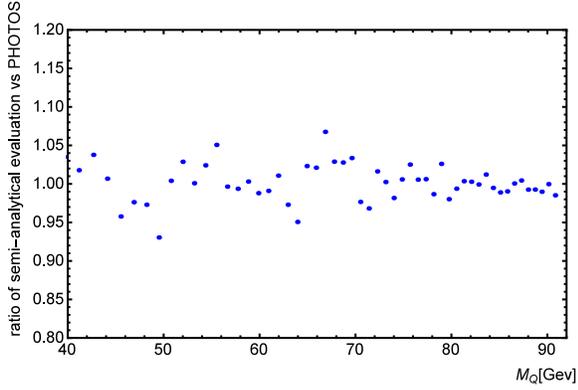}} a)  $p p\rightarrow Z\rightarrow e^{+}e^{-}(e^{+}e^{-})$\\
\end{minipage}
\hfill
\begin{minipage}[h]{0.47\linewidth}
\center{\includegraphics[width=1\linewidth]{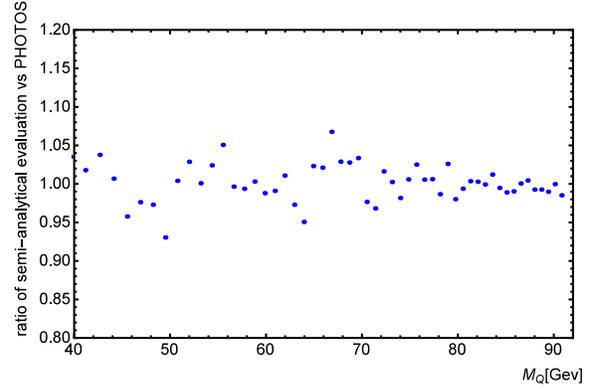}} \\b) $p p\rightarrow Z\rightarrow e^{+}e^{-}(\mu^{+}\mu^{-})$
\end{minipage}
\vfill
\begin{minipage}[h]{0.47\linewidth}
\center{\includegraphics[width=1\linewidth]{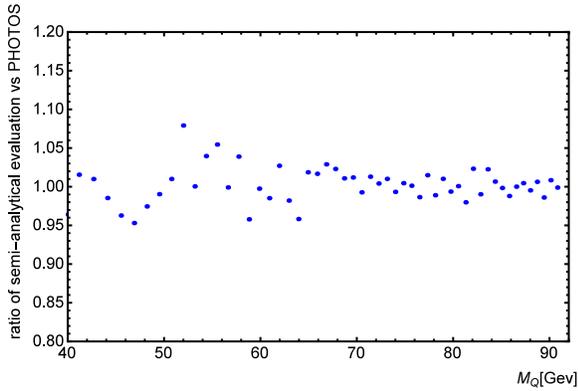}} c) $p p\rightarrow Z\rightarrow \mu^{+}\mu^{-}(e^{+}e^{-}) $\\
\end{minipage}
\hfill
\begin{minipage}[h]{0.47\linewidth}
\center{\includegraphics[width=1\linewidth]{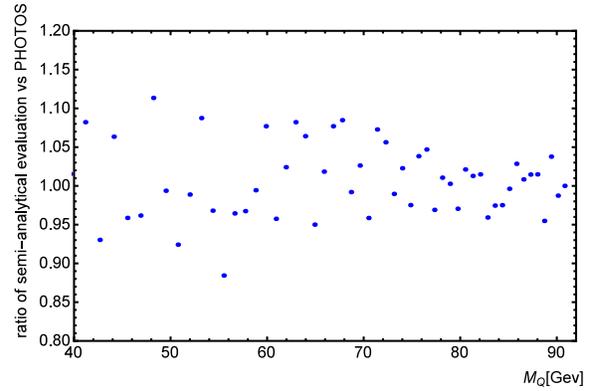}} d) $p p\rightarrow Z\rightarrow \mu^{+}\mu^{-}(\mu^{+}\mu^{-})$\\
\end{minipage}
\end{minipage}
\end{center}
\caption[] {Number of events from {\tt PYTHIA} multiplied by a factor resulting from formula~(\ref{BFactor}) divided by number of events from {\tt PYTHIA}$\times${\tt PHOTOS}. For these particular plots there is difference in {\tt PYTHIA} initialization parameters; energy range of leptonic system is limited to $[91.183$$,91.252$$]$ GeV window.
\label{fig:narrow_peak}}
\end{figure}

\begin{figure}[htp!]
\begin{center}
\begin{minipage}[t]{1.0\textwidth}
\begin{minipage}[h]{0.47\linewidth}
\center{
\small{
\begin{verbatim}
WeakSingleBoson:ffbar2gmZ = on
23:onMode = off
23:onIfAny = 11
23:mMin    = 10.0
23:mMax    = 200.0
HadronLevel:Hadronize = off
SpaceShower:QEDshowerByL = off
SpaceShower:QEDshowerByQ = off
PartonLevel:ISR = off
PartonLevel:FSR = off
Beams:idA =   2212
Beams:idB =   2212
Beams:eCM =  14000.0
\end{verbatim}
}
}
\flushleft a)  $p p\rightarrow Z\rightarrow e^{+}e^{-}(e^{+}e^{-},\mu^{+}\mu^{-})$
\end{minipage}
\hfill
\begin{minipage}[h]{0.47\linewidth}
\center{
\small{
\begin{verbatim}
WeakSingleBoson:ffbar2gmZ = on
23:onMode = off
23:onIfAny = 13
23:mMin    = 10.0
23:mMax    = 200.0
HadronLevel:Hadronize = off
SpaceShower:QEDshowerByL = off
SpaceShower:QEDshowerByQ = off
PartonLevel:ISR = off
PartonLevel:FSR = off
Beams:idA =   2212
Beams:idB =   2212
Beams:eCM =  14000.0
\end{verbatim}
}
}
\flushleft b) $p p\rightarrow Z\rightarrow \mu^{+}\mu^{-}(e^{+}e^{-},\mu^{+}\mu^{-})$
\end{minipage}
\end{minipage}
\end{center}
\caption[] {Initialization parameters for {\tt PYTHIA}.
\label{fig:initialization}}
\end{figure}

\end{document}